\documentclass[preprint2,eqsecnum]{aastex}
\usepackage{amssymb,amsbsy,graphics}

\begin{document}
\newlength{\colw}
\setlength{\colw}{3.2in}

\title{Fragmentation Instability of Molecular Clouds: 
Numerical Simulations\altaffilmark{1}}
\author{R\'{e}my Indebetouw\altaffilmark{2}}
\affil{CASA, University of Colorado, Boulder, CO, 80309}
\and
\author{Ellen G. Zweibel\altaffilmark{3}}
\affil{JILA, University of Colorado, Boulder, CO, 80309}
\altaffiltext{1}{in press, ApJ 532, April 1, 2000}
\altaffiltext{2}{indebeto@casa.colorado.edu}
\altaffiltext{3}{zweibel@solarz.colorado.edu}

\begin{abstract}
We simulate fragmentation and gravitational collapse of cold, 
magnetized molecular
clouds.  We explore the nonlinear development of an 
instability mediated by ambipolar 
diffusion, in which the collapse rate is intermediate to 
fast gravitational collapse and slow quasistatic collapse. 
Initially uniform stable clouds fragment into elongated
clumps with masses largely determined by the cloud 
temperature, but substantially larger than the thermal Jeans mass.  
The clumps are asymmetric, with significant 
rotation and vorticity, and lose magnetic flux as they collapse.
The clump shapes, intermediate collapse rates, and infall profiles 
may help explain observations 
not easily fit by contemporary slow or rapid collapse
models.
\end{abstract}

\section{Introduction}

The interstellar magnetic field 
plays an important role in the dynamics of molecular clouds 
and the collapse of dense cloud cores into protostars.  
Magnetic pressure and tension combine with thermal and turbulent 
kinetic pressure to resist gravitational collapse.  The role of 
the magnetic field is 
often simply characterized by a critical mass $M_{\rm crit}$
which depends on the magnetic flux threading the cloud 
\citep{ms56,ms76,tin88,mckee93,m99}.
Clouds with $M > M_{\rm crit}$ are termed
supercritical and collapse on a dynamical time-scale: 
the magnetic field can have a moderate effect on the morphology of 
collapse and can slow collapse in the cloud envelope
\citep[{\it e.g.} ][]{black82}, but cannot significantly slow collapse in 
the core.  
Clouds with $M < M_{\rm crit}$ are termed subcritical, and
evolve on a longer 
time-scale as magnetic support is lost due to ambipolar diffusion. 
The magnetic field is redistributed within the cloud
so that the inner parts become supercritical. 
The cloud is then differentiated
into a dynamically collapsing core with a magnetically supported
envelope \citep{cm93,fm93,bm94,cm94,sms97,ck98}.
Much progress has been made in following this
type of evolution through 6 or more orders of magnitude of increase in
central density, including the effects of rotation as well as detailed
chemistry and grain physics. The outcomes of these calculations include
detailed density, bulk velocity, ion-neutral drift velocity, 
magnetic field, and grain and
ion abundance profiles in axisymmetric clouds, as well as an appreciation of
the timescales, rate of magnetic flux loss, and role of magnetic braking in
this mode of isolated star formation. 

This theoretical picture can be
observationally tested \citep[see recent reviews by ][]{evans99,meo99}.
So far, the results are ambiguous. Flow toward an isolated infrared source 
in the Bok globule B335 is well described by an inside-out collapse model 
\citep{z90,z93,z95}.
The density distribution and measured magnetic
fieldstrength in the cloud B1 have been fit by a model with a subcritical
envelope and a core which has evolved to a supercritical state by ambipolar
drift \citep{c94}. On the other hand, in some respects the existing models
appear to be incomplete. Clumps and cores are not axisymmetric; \citet{m91} 
surveyed 16 dense
cores a few tenths of a parsec in size and pointed out that at least 6 of
them are likely to be prolate. \citet{r96} made a statistical 
argument, based on a larger sample, that clumps and globules are more likely
prolate or triaxial than
oblate. \citet{wt99} showed that asphericity appears also at 
smaller scales. Collapse guided by a magnetic field could produce oblate
clouds, but not prolate ones. 
It also appears unlikely that rotation accounts for the flattening;
this has been shown quantitatively in the case of L1527 \citep{o97}.
Thus, the shapes are unexplained. Moreover,
in some cases in which infall has been measured directly, it is more spatially
extended, with faster velocities in the outer parts, than expected from
the standard models of inside-out collapse or gravitational motion driven by
ambipolar drift. This has been shown in the case of L1544 by \citet{t98}, and
for 6 other starless cloud cores by \citet{greg}. 
%rev8
(Interestingly, a model of L1544 has been recently constructed by 
\citet{cb2000} to match the observations of \citet{t98}.  The model 
requires a mass-to-flux ratio more nearly critical than previously published 
models by the same authors, and appears to display the same intermediate
collapse discussed in this paper, although the authors do not call it out as 
such.)
%rev8
Finally, there are observations which relate to the timescale for collapse 
of molecular clouds into protostars.  \citet{lm99} find 
collapse timescales of $\sim$ 0.3-1.6 Myr from the ratio of the 
numbers of starless cores to cores with embedded young stellar 
objects.  They state that this requires collapse 2-44 times 
faster than ambipolar drift models.
These observations suggest that another ingredient may be required
to explain the collapse of molecular clouds: the decay of turbulent 
support \citep{mlazar}, or, as we explore in this
paper, a magnetogravitational instability mediated by ambipolar drift.

The magnetic field likely also plays an important role in other
aspects of collapse, which are still incompletely understood.
The mechanism by which molecular clouds fragment,
and the masses and morphology of those fragments, is clearly of
importance to the stellar initial mass function and the origin of
binary systems.  The magnetic
field can exert strong forces on many scales, affecting fragmentation
\citep[{\it e.g.} ][]{boss97,boss99}. 
The field may also be a source of kinetic energy in the cloud,
if the free energy of an ordered field can be released as turbulent
motions. 
These issues have not been studied 
thoroughly in the detailed
axisymmetric  models of gravitational contraction with
ambipolar drift referenced above because they would require
calculations with no constraints on spatial symmetry.

In this paper, we address some of these issues by studying a simple problem:
the evolution of small perturbations to an initially uniform, 
magnetically subcritical sheet of
weakly ionized gas with a
uniform magnetic field perpendicular to
its plane. The perturbations evolve under the influence of magnetic tension,
self gravity, thermal pressure, and ambipolar drift. Typically the sheet
breaks up into a small number of fragments of elongated shape which are
collapsing, losing magnetic flux through ambipolar drift, and
interacting gravitationally with one another. The magnetic
fields associated with these asymmetric clumps generate local vorticity (the
net angular momentum of the sheet is identically zero). The characteristic
vorticity structure is a vortex pair which flanks each clump and is
associated with strong streaming motions along it. The clump masses are
typically of order 1-10 $M_{\odot}$, and scale with temperature like the Jeans
mass, but are typically larger because of magnetic support. 
The main features of collapse in this geometry were predicted 
by the linear stability analysis of Zweibel (1998; hereafter Z98); there is
also some overlap with the earlier stability analysis in 3D by 
\citet{langer78}. 
Here we follow the evolution into the nonlinear regime and follow the growth
of density fluctuations from .01 to up to 10 times the mean surface 
density.
Collapse occurs on an intermediate time-scale, slower than the dynamical 
or free-fall time-scale, but faster than the ambipolar diffusion 
time-scale.  As noted above, 
this intermediate collapse rate may be observed in 
some clouds \citep[{\it e.g.} ][]{lm99,greg} .
As no restrictions are placed on the 
morphology of the clumps, we can begin to explore
the nature of flux loss and collapse in more complicated geometries 
than isolated axisymmetric clouds. One outcome suggested by linear
theory which we have not resolved 
is whether stored magnetic energy is converted to turbulence, as there
is no stored magnetic energy in the system initially. 

The unperturbed initial geometry, governing equations, and linear 
theory are described in \S\ref{setup}.  Section 
\ref{num} contains a 
description of the numerical method and main results:  
the collapse rate is discussed in \S\ref{gamgam}, 
the relationship between magnetic field {\bf B} and density 
$\rho$ or surface density
$\sigma$ in \S\ref{brho}, size of fragments in \S\ref{clump}, 
velocity structure in \S\ref{vel}, 
and distribution of energy in \S\ref{energy}.
The validity of the approximations is discussed in \S\ref{discuss}, 
and the summary and conclusions are in 
\S\ref{summary}.

%%%%%%%%%%%%%%%%%%%%%%%%%%%%%%%%%%%%%%%%%%%%%%%%%%%%%%%%%%%%%%%%%%%%%%%
\section{Governing Equations and Linear Theory}\label{setup}

We simulate a flat slab of cold gas with the 
magnetic field initially 
perpendicular to the slab (the $\hat{z}$ direction).  
Z98 discusses the linear theory for 
this model when the cloud temperature T equals 0.  
In this section we review the model in the more 
general case T$>$0.  We recall the results of the linear theory 
and discuss the consequences of adding thermal pressure.

A flat slab-like model has observational and theoretical motivation: 
molecular clouds commonly have sheet or filament-like 
structure (although detailed, high-resolution 
information on the field orientation in such structures 
is not yet available for many objects).  In a fairly quiescent environment, a
roughly spherical molecular cloud with a large-scale, 
dynamically significant,
ordered magnetic field will relax into a pancake or slab
as matter drains down the field lines.  Magnetic forces 
will allow comparatively little contraction perpendicular 
to the field direction, resulting in a slab with a predominantly 
perpendicular field. Such slabs could also be formed by shock waves 
propagating parallel to the local magnetic field. 

We use a simple initial state with small ($<$1\%) 
perturbations.  The boundary conditions are periodic in the horizontal 
($\hat{x}$ and $\hat{y}$) directions and the initial surface 
density $\sigma_0$ and magnetic field $B_{z0}$ are uniform.  
%rev8
The unperturbed initial state was chosen for computational 
simplicity, as self-consistent finite disk and slab-like equilibrium 
states cannot generally be described by simple expressions in closed form.
\citep[{\it e.g.} ][]{park74,mous76,bau89,mestel85}.
We note that our initial state is not technically one of static equilibrium, 
but rather a version of the commonly used Jeans swindle, in which 
the unperturbed gravitational potential is discarded
(see {\it e.g.} discussion in \citet{bt}).

%%%%%%%%%%%%%%%%%%%%%%%%%%%%%%%%%%%%%%%%%%%%%%%%%%%%%%%%%%%%%%%%%%%%%%%%
\subsection{Governing Equations}

We begin with the equations of ideal magnetohydrodynamics 
for two inviscid, non-resistive, interacting, magnetized fluids, 
one charged and the other neutral.  Since sources and sinks are 
expected to dominate advection in the ion continuity equation, 
we treat directly only the continuity equation for neutrals, 
and parameterize the ion behavior through equation 
(\ref{eqn_alpha}).
The governing MHD equations are thus the equation of continuity
for neutral particles,
\begin{mathletters}
\begin{equation} 
{{\partial\rho_n}\over{\partial t}} + 
	\boldsymbol\nabla\cdot\left(\rho_n\mathbf{v}\right) = 0,
\end{equation}
equations of motion for the two species, 
\begin{eqnarray}
\rho_n{{ D_n\mathbf{v}_n}\over{ D_nt}}
	+ \boldsymbol\nabla P_n 
	+ \rho_n\boldsymbol\nabla\Phi_G \\\nonumber
	+ \rho_n\nu_{ni}(\mathbf{v}_n-\mathbf{v}_i) &=& 0, \\
\rho_i{{ D_i\mathbf{v}_i}\over{ D_it}}
	+ \boldsymbol\nabla P_i 
	+ \rho_i\boldsymbol\nabla\Phi_G \\\nonumber
	- \rho_i\nu_{in}(\mathbf{v}_n-\mathbf{v}_i) &=& 
	{{(\boldsymbol\nabla\times\mathbf{B})\times\mathbf{B}}
	\over{ 4\pi}}, 
\end{eqnarray}
the induction equation
\begin{equation}
{{\partial\mathbf{B}}\over{\partial t}} 
= \boldsymbol\nabla\times\left(\mathbf{v}_i\times\mathbf{B}\right),
\end{equation}
and Poisson's equation
\begin{equation}
 \nabla^2\Phi_G = 4\pi G\rho. 
\end{equation}
\end{mathletters}

Subscripts $i$ and $n$ denote ions and neutral particles respectively.
$D_{\alpha}/D_{\alpha}t$ is the convective derivative for species $\alpha$.
The ion-neutral collision frequency is 
$\nu_{in} = \rho_n \langle \sigma v \rangle/(m_i + m_n)$,
and  $\nu_{ni}$ is the neutral-ion 
collision frequency $(\rho_i \nu_{in} = \rho_n \nu_{ni})$. In the context
of collision rates only, the
symbol $\sigma$ represents the cross-section for elastic collisions; elsewhere
it represents surface density.
The gravitational potential is $\Phi_G$, and $\mathbf{v}$, $\rho$, $P$, and 
$\mathbf{B}$ are the 
velocity, density, pressure, and magnetic field, respectively.
We work on large scales and at low temporal frequencies for which the 
ions and electrons are coupled. 

We assume that the ionization fraction in the cloud is low. For
dense molecular gas which is ionized by cosmic rays and recombines on grains,
$n_i/n_n \sim K n_n^{-1/2}$, where $K\sim 1.1\times 10^{-5}$ \citep{mckee93}
Departures from and generalizations of this ionization law are discussed
below, see eq. [\ref{eqn_alpha}].
Therefore, 
$\rho_n \approx \rho$ and $\mathbf{v}_n \approx \mathbf{v}$. If the neutral-ion
collision time is much less than a dynamical time, 
the ambipolar drift velocity 
$\mathbf{v}_D$ can be written in the standard form \citep{shu83}:
\begin{equation}
\mathbf{v}_D\equiv\mathbf{v}_i-\mathbf{v}_n = 
{{(\boldsymbol\nabla\times\mathbf{B})\times\mathbf{B}}
\over{4\pi\nu_{in}\rho_i}}.
\end{equation}

The flat geometry allows significant simplification of the equations
by taking a vertical integral in the limit of 
infinitesimal vertical thickness.  For example, the 
magnetic force simplifies to:
\begin{eqnarray}
\lim_{\epsilon\rightarrow 0}\int\limits_{-\epsilon}^{+\epsilon} dz
{{(\boldsymbol\nabla\times\mathbf{B})\times\mathbf{B}}
\over{ 4\pi}} &=& \\
\lim_{\epsilon\rightarrow 0}{{ B_z}\over{ 4\pi}}
\left[B_x\hat{x} + B_y\hat{y}\right]_{-\epsilon}^{+\epsilon} &=&
{{ B_z\mathbf{B}_h}\over{ 2\pi}}.
\end{eqnarray}
In the limit of an infinitesimally thin disc or slab, 
the vertical component of the magnetic field $B_z$ is 
continuous with
respect to the plane of the slab (the $\hat{z}$ direction), and 
the horizontal component $\mathbf{B}_h$ is 
antisymmetric (reverses sign) with respect to the plane of the slab.

In addition, we assume the vertical component of the velocity is 
negligible compared to the
horizontal components $(v_z \ll v_x, v_y)$.  \citet{lz97} found that 
thin disks are generally stable to warping modes, so we expect 
predominantly horizontal motion.

In the limit of zero gas density and thermal pressure outside the 
slab, the external magnetic field relaxes instantaneously to 
an equilibrium state, shown in Z98 to be a current-free or 
potential field state.
We therefore assume that the magnetic field at $z \neq 0$ is a potential field.
This allows us to calculate only the 
vertical part of the magnetic field in the disc, rather than all 
three components in a three-dimensional domain, a
tremendous simplification.  Limitations of this 
approximation, and corrections
to it, are discussed in more detail in \S\ref{potfield}
and the Appendix.

The resulting system of 2-dimensional equations are as follows:
The equation of continuity 
\begin{mathletters}
\begin{equation}
\label{gov1}
{{\partial\sigma}\over{\partial t}}
	+\boldsymbol\nabla_h\cdot(\sigma\mathbf{v}_h) = 0,
\end{equation}
of motion
\begin{eqnarray}
\label{gov2}
\sigma{{ D\mathbf{v}_h}\over{ D t}}
	+ \boldsymbol\nabla_h P
	+ \sigma\boldsymbol\nabla_h\Phi_G &=& 
	{{\mathbf{B}_h B_z}\over{ 2\pi}}, \\
v_z &=& 0,
\end{eqnarray}
the definition of the ambipolar drift velocity
\begin{equation}
\mathbf{v}_{Dh} =
	{{ B_z\mathbf{B}_h}\over
	{ 2 \pi\nu_{in}\sigma_i}},
\end{equation}
the induction equation
\begin{equation}
{{\partial B_z}\over{\partial t}} =
	-\boldsymbol\nabla_h\cdot\left[(\mathbf{v}_h+\mathbf{v}_{Dh})
	B_z\right], 
\end{equation}
the potential field equations
\begin{eqnarray}
B_z &=& {{\partial\Phi_M}\over{\partial z}}, \\
\mathbf{B}_h &=& \mathbf\nabla_h\Phi_M, 
\end{eqnarray}
and Poisson's equation
\begin{equation}
\label{gov}
{{\partial\Phi_G}\over{\partial z}}=2\pi G\sigma.
\end{equation}
\end{mathletters}
The gravitational and magnetic potentials are $\Phi_G$ and $\Phi_M$ 
respectively, 
$\sigma$ is the surface density, and $h$ denotes horizontal components 
({\it e.g.}, $\mathbf{v}_{Dh}$ is the horizontal drift velocity).
We assume an isothermal equation of state
\begin{math}P = a^2\sigma \end{math}.  

Equations for an
axisymmetric disk of small but finite half thickness $Z$ were derived by
\citet{cm93}. Their equations contain correction terms of order $Z/R$, where
$R$ is the distance from the axis of symmetry; these terms include magnetic
pressure, which provides a restoring force which we neglect, and corrections
to the normal direction, which is tilted slightly from the vertical because
$Z$ depends on $R$. These terms go smoothly to zero in the limit 
$Z/R\rightarrow 0$.  For canonical values of physical parameters in dense
molecular clouds, we find $Z/R < 1/10$. Strictly speaking, we should retain
the magnetic pressure gradient, as \citet{cm93} do, because it is of the
same order as, although generally less than, the thermal pressure gradient,
and also provides a restoring force. 
However, magnetic tension clearly dominates magnetic pressure at 
the long wavelengths of greatest interest here, while as the wavenumber 
increases the rate of ambipolar drift increases as well, so that the 
magnetic pressure force at short wavelengths is less than it would be 
if the magnetic field were frozen in. Moreover, we know that there
is a 3D version of the instability which is driven by magnetic pressure 
alone.  We study the instability driven by tension, but the existence 
and nature of the instability should be the same whether it is driven 
by tension or pressure.
For all these reasons, we think that the neglect of magnetic 
pressure is not a major source of error.

%%%%%%%%%%%%%%%%%%%%%%%%%%%%%%%%%%%%%%%%%%%%%%%%%%%%%%%%%%%%%%%%%%%%%%%%
\subsection{Nondimensionalization}
\label{nondim}

All quantities in the problem are scaled by a self-consistent set of 
characteristic quantities.  
Given an initial vertical magnetic field
$B_{z0}$, we choose as the characteristic surface density that which is 
marginally stable 
to collapse in the zero temperature limit \citep{nn78},
\begin{equation}
\label{eqn_critical}
 \sigma_{c0} \equiv
	{{ B_{z0}}\over{ 2\pi G^{1/2}}}.
\end{equation} 
In a 3-dimensional model 
of a cold magnetized molecular cloud, one logical choice would be to
use the Alfv\'{e}n velocity as a characteristic velocity.
The 2-dimensional geometry precludes this approach, because the 
quantity $B_z /\sqrt{2\pi\sigma}$ which arises naturally has dimensions not
of (a 2-D Alfv\'{e}n) velocity but rather of 
length\textsuperscript{1/2}/time.  A length
scale is thus required.  A logical choice in this geometry is the scale height
of the slab, $H = a^2 / 2 \pi \sigma G$, but this is undesirable because 
the problem of most interest is a cold cloud, in which the isothermal sound
speed $a^2 \rightarrow 0$. 
Instead we choose a characteristic length scale $L$, which
will be the horizontal domain size, or equivalently the largest spatial
wavelength in the simulation (Of course, $L$ scales out of all final results
when expressed in dimensional units).  
The characteristic velocity is the 
Alfv\'{e}n speed for the critical surface density and magnetic field
\begin{equation}
\label{eqn_alfven}
 v_{a0} \equiv B_{z0}\sqrt{{ L}
	\over{ 2\pi\sigma_{c0}}}, 
\end{equation}
and the characteristic time is simply
\begin{equation}
\label{eqn_time}
 t_{c0} \equiv {{ L}\over{ v_{a0}}}. 
\end{equation}
The nondimensionalized variables are as follows:
\begin{equation}
\begin{array}{rclrcl}
\omega &\equiv& {{\sigma}\over{\sigma_{c0}}}, &
	\boldsymbol\beta &\equiv&
	{{\mathbf{B}}\over{ B_{z0}}}, \\
\tau &\equiv& {{ t}\over{ t_{c0}}}, &
	(\mu,\nu) \;=\; \boldsymbol\nu &\equiv&
	{{\mathbf{v}}\over{ v_{a0}}}, \\
\nabla_h &\leftarrow&
	{{\nabla_h}\over{ L}}, &
	\xi,\eta,\zeta &\equiv& 
	{{ x,y,z}\over{ L}}, \\
\phi_G &\equiv& 
	{{ t_{c0}^2}\over{ L^2}}\Phi_G, & \mbox{and}\hspace{3ex}
	\phi_M &\equiv& 
	{{\Phi_M}\over{ B_{z0}L}}.
\end{array}
\end{equation}

%%%%%%%%%%%%%%%%%%%%%%%%%%%%%%%%%%%%%%%%%%%%%%%%%%%%%%%%%%%%%%%%%%%%%%%%
\subsection{Parameterization of Ambipolar Drift}

We assume that the product of the ion surface density $\sigma_i$
and the ion-neutral collision frequency $\nu_{in}$ is 
related to the surface density $\sigma\simeq\sigma_n$ according to the simple
{\it ansatz}:
\begin{equation}
\label{eqn_alpha}
{{\delta(\sigma_i\nu_{in})}\over
	{\sigma_i\nu_{in}}} =
	\alpha{{\delta\sigma}\over{\sigma}}.
\end{equation}
%%%
If the ionization fraction $x$ scales as the
neutral density $n_n^{-q}$, and the scale height $H$ scales as $\sigma^{-1}$,
then $\alpha=3-2q$. Often $q$ is taken to be 0.5 
\citep{mckee93}, but a detailed treatment
of grain dynamics in contracting cores \citep{cm94,cm95,cm98}
shows that the parameter $q$ continuously decreases throughout contraction, 
and may
range from $\sim 0.6$ to less than $0.1$ as the central density increases by
about 6 orders of magnitude.
We have tested the sensitivity of our
results to the value of $\alpha$ by comparing models with $\alpha$
 ranging from
0 to 3, and find that the results differ by less than 5\% as the density
perturbations grow from .01 to 1.5 times the
mean density (This is consistent with linear 
perturbation theory, which predicts that the value of $\alpha$ enters only if
%rev8
the unperturbed initial state has an inclined magnetic field; Z98). In view of
the insensitivity of the results to $\alpha$, as well as the fact that the
volume density  increases by less than 2 orders of magnitude in our
calculations, we regard the {\it ansatz} 
equation (\ref{eqn_alpha}) as adequate.

To compare with the linear theory, we use a nondimensional form of 
the drift frequency 
$\Gamma = t_{c0} k B_{z0}^2 / 2\pi\sigma_i\nu_{in}$.
(Z98 uses the {\it dimensional} 
$\Gamma = k B_{z0}^2 / 2\pi\sigma_i\nu_{in}$.)
In the simulation, it is convenient to compute the evolution of 
the drift independent of spatial wavenumber $k$.  We use the 
quantity 
$\Gamma/kL$, which is $\Gamma/2\pi$ for the lowest spatial 
wavenumber $k=2\pi/L$.

The full set of governing equations in conservative form are 
as follows:
\begin{mathletters}
\begin{eqnarray}
{{\partial\omega}\over{\partial\tau}} 
	&=& -\boldsymbol\nabla_h\cdot(\omega\boldsymbol\nu_h), \\
{{\partial}\over{\partial\tau}}\omega\boldsymbol\nu_h &=&
	-\boldsymbol\nabla_h\cdot(\omega\boldsymbol\nu_h\boldsymbol\nu_h)
	-a^2\boldsymbol\nabla_h\omega \nonumber\\&&
	-\omega\boldsymbol\nabla_h\phi_G
	+\beta_z\boldsymbol\beta_h, \\
\omega &=& \left.{{\partial\phi_G}\over
	{\partial\zeta}}\right|^+, \\
\boldsymbol\beta_h &=& \boldsymbol\nabla_h\phi_M, \\
	\beta_z &=& 
	{{\partial\phi_M}\over{\partial\zeta}}, \\
{{\partial\beta_z}\over{\partial\tau}} &=& 
	-\boldsymbol\nabla_h\cdot\left[\beta_z(\boldsymbol\nu
	+\boldsymbol\nu_D)\right], \\
\boldsymbol\nu_D &=& \beta_z\boldsymbol\beta_h
	{{\Gamma}\over{ kL}}, \\
{{\partial}\over{\partial\tau}}
\left({{ kL}\over{\Gamma}}\right) &=&
\label{eqn_drift}
	{{ kL}\over{\Gamma}}
	{{\alpha}\over{\omega}}
	{{\partial\omega}\over{\partial\tau}}.
\end{eqnarray}
\end{mathletters}

Note that we interpret
equation (\ref{eqn_alpha})
as an Eulerian relation in equation (\ref{eqn_drift}).

%%%%%%%%%%%%%%%%%%%%%%%%%%%%%%%%%%%%%%%%%%%%%%%%%%%%%%%%%%%%%%5
\subsection{Physically Reasonable Parameter Regime}
\label{phys_par}

The input parameters for the model are the sound speed $a$, 
the strength of the initial magnetic field relative to the 
surface density $B_{z0}/2\pi G^{1/2}\sigma_0 = 1/\omega_0$, 
the drift parameter $\Gamma$, and the constant $\alpha$
which determines the perturbation to the collision
rate (see eq. [\ref{eqn_alpha}]).

Typical magnetic fields in dense clouds are 30$\mu$G \citep{88.329}, 
and we choose the nondimensionalization length scale
(see \S\ref{nondim}) to be 1pc, a typical 
size for a dense cloud or cloud core and its 
close neighborhood.  A typical cold 
cloud temperature is 10K, and the average molecular weight is 
that of molecular hydrogen with 10\% helium, $m_n$ = 
3.9$\times$10\textsuperscript{-24}g. 
This yields the following expressions for the Alfv\'{e}n and 
sound speeds:
\begin{mathletters}
\begin{eqnarray}
v_{a0}^2 &=& B_{z0}^2{L\over{2\pi\sigma_{c0}}} \\
	&=& B_{z0}LG^{1/2} \nonumber\\
	&=& 2.4\times 10^{10} {{{\rm cm}^2}\over{{\rm s}^2}}
		\left({B_{z0}\over{30\mu {\rm G}}}\right)
		\left({L\over{\rm pc}}\right), \nonumber\\
a^2 &=& {{k_BT}\over{m},} \\
{{ a^2}\over{ v_{a0}^2}} 
	&\simeq& 0.02 \left({T\over{10 {\rm K}}}\right)
	\left({{\rm pc} \over L}\right)
	\left({{30\mu {\rm G}}\over{B_{z0}}}\right) \nonumber\\&&\times
	\left({{3.9\times 10^{-24}{\rm g}}\over{m_n}}\right).
\end{eqnarray}
\end{mathletters}
From now on, $a^2$ will be given in units of $v_{a0}^2$.

The initial surface density is chosen to be
$\simeq$ 0.02 g cm\textsuperscript{-2}, or about 100 M$_{\odot}$ pc$^{-2}$. 
This corresponds not only to a typical column density for a dense
cold cloud ($N_H\sim 10^{22}$ cm$^{-2}$), but also to the surface 
density that would result if 
a spherical cloud of typical number density 
(10\textsuperscript4 cm\textsuperscript{-3}) 
and typical size (several 10\textsuperscript{18}cm) collapsed 
along a large-scale magnetic field into a thin pancake or disc. 
The following expression for the normalized surface density 
results:
\begin{eqnarray}
\label{eqn_omega_0}
\omega_0 &=& {{2\pi G^{1/2}\sigma_0}\over{B_{z0}}} \\
	&\simeq& 1.0 \left({{30\mu {\rm G}}\over{B_{z0}}}\right)
	\left({{\sigma_0}\over{0.02\ {\rm g}\ {\rm cm}^{-2}}}\right).\nonumber
\end{eqnarray}
Thermal pressure raises the critical surface 
density for gravitational collapse, and for a cloud at
T = 10K the value of $\omega_0$ for the fiducial parameters in equation 
(\ref{eqn_omega_0})
is 0.864 of that critical surface density $\omega_{\rm crit}$.
%%%

Thermal pressure imparts a finite scale height to the disk 
\begin{eqnarray}
\label{eqn_H}
H &=& {{a^2}\over{2\pi\sigma_0 G}} \\
        &\simeq& 4\times10^{16}{\rm cm} \left({T\over{10 {\rm K}}}\right)
                \left({{0.02\ {\rm g}\ {\rm cm}^{-2}}\over{\sigma_0}}\right) \nonumber\\&&\times
                \left({{3.9\times 10^{-24} {\rm g}}\over{m_n}}\right).\nonumber
\end{eqnarray}

Clearly, $H\ll L$. An inclined magnetic field exerts a pinching force,
compressing the disk and reducing $H$ further \citep{wardle93}.

Determination of the drift parameter $\Gamma$ requires an expression 
for the neutral-ion collision frequency, 
$\nu_{ni}\simeq2\times 10^{-9} n_i$ cm\textsuperscript{3} 
s\textsuperscript{-1} 
\citep{draine83} and the density of ions. We assume
$n_i = Kn_n^{1/2}$, where $K 
\simeq 1.1\times 10^{-5}\ n_n^{1/2}$ cm\textsuperscript{-3/2}
\citep{mckee93}. Using these relations
\begin{eqnarray}
\label{eqn_Gamma}
\Gamma &=& {k t_{c0}{B_{z0}^2}\over{2\pi\sigma_i\nu_{in}}} \\
	&=& {k t_{c0}{B_{z0}^2}\over{2\pi\sigma_0\nu_{ni}}} \nonumber\\
	&\sim& 0.059 \left({{\rm pc}\over L}\right)^{1\over 2}
		\left({{B_{z0}}\over{30\mu {\rm G}}}\right)^{3\over 2}
		\left({{0.02\ {\rm g}\ {\rm cm}^{-2}}\over{\sigma_0}}\right) \nonumber\\&&\times
		\left({{5\times 10^4\ 
                  {\rm cm}^{-3}}\over{n_n}}\right)^{1\over 2}
		\left({k\over{2\pi/L}}\right).\nonumber
\end{eqnarray}
The fiducial value of $n_n$ which appears here is consistent with the other
parameters: $n_n = \pi G\sigma_0^2/k_BT$.

Although equation (\ref{eqn_Gamma}) shows that $\Gamma$ does
not depend on the scale height $H$, since the temperature of molecular clouds
is quite well determined it is useful to rewrite $\Gamma$ in a way
which does depend on $H$ and suppresses the dependence on some of the other
parameters. We have
\begin{eqnarray}
\label{eqn_Gamma_alt}
\Gamma &=& {{4\pi^{3/2}(m_n G)^{3/2}}\over {K\langle\sigma v\rangle\omega_0^{
3/2}}}\Bigg({{H}\over {L}}\Bigg)^{1/2}\\
&=&0.56\Bigg({{H}\over {L}}\Bigg)^{1/2},\nonumber
\end{eqnarray}
where we have used the standard values for all the constants. Equations
(\ref{eqn_H}) and 
(\ref{eqn_Gamma_alt}) suggest $\Gamma\le 0.1$ for typical parameters.

%%%%%%%%%%%%%%%%%%%%%%%%%%%%%%%%%%%%%%%%%%%%%%%%%%%%%%%%%%%%%%%%%%%%%%%%
\subsection{Linear Theory}
\label{lin_th}

The collapse rate for a linear perturbation with positive thermal pressure 
can be easily calculated (see Z98 for the calculation at T=0K).
The physical quantities $\omega$, $\beta_z$, $\boldsymbol\nu$
and governing equations 
(\ref{gov1}-\ref{gov}) are linearized and reduced to one horizontal spatial 
dimension. Assuming a single Fourier mode
\begin{mathletters}\begin{eqnarray}
\omega &\rightarrow& \omega_0 + \omega e^{\gamma\tau+ik\xi-k|\zeta|}, \\
\beta_z &\rightarrow& 1 + \beta_z e^{\gamma\tau+ik\xi-k|\zeta|}, \\
\nu &\rightarrow& \nu e^{\gamma\tau+ik\xi-k|\zeta|}
\end{eqnarray}\end{mathletters}
leads to the dispersion relation 
\begin{equation}
\gamma^3 + \Gamma\gamma^2 + 
\left[\gamma_G^2\left({1\over{\omega_0^2}}-1\right)+\gamma_T^2\right]\gamma
	- \gamma_G^2\Gamma + \gamma_T^2\Gamma = 0, 
\label{disp_rel}
\end{equation}
where $\gamma_G = \sqrt{\omega_0kL}$ is the nondimensionalized gravitational 
frequency, and $\gamma_T = akL$ is the nondimensionalized 
thermal frequency.
%%%

Equation (\ref{disp_rel}) has two limits which provide some insight into what
follows. In the limit $\Gamma = 0$ we recover the dispersion relation for
waves driven by magnetic tension, thermal pressure, and self-gravity; the
first two forces are stabilizing and the last is destabilizing. The system is
stable for all wavenumbers if $\omega_0 < 1$, but if $\omega_0 > 1$, the
system is unstable for wavenumbers $k < k_c(\Gamma=0)\equiv 
 H^{-1}(\omega_0^2-1)/\omega_0^2$. (In
a system of finite size $L$, $k$ is bounded from below by $2\pi/L$, leading to
the absolute stability criterion $\omega_0 < \omega_{crit}$ which we present
below).
 The maximum growth rate, in dimensional form
(recall that in equation (\ref{disp_rel}), 
$\gamma$ is given in units of $t_{c0}^{-1}$
), is $\gamma_{max}(\Gamma=0)=\pi G\sigma_0(\omega_0^2-1)/(a\omega_0^2)$, and
occurs at a wavenumber $k_m(\Gamma = 0) = {{1}\over {2}}k_c(\Gamma=0)$ (in
these dimensional expressions, $a$ is the {\it dimensional} sound speed).

In the limit of large $\Gamma$, the magnetic field is uncoupled from the gas,
and the dispersion relation reverts to that of an unmagnetized slab. The
system is unstable for $\gamma_G^2 > \gamma_T^2$, or $k < 
k_c(\Gamma\rightarrow\infty) \equiv 2\pi G\sigma_0/a^2
= H^{-1}$. The maximum growth rate, which occurs at 
$k_m(\Gamma\rightarrow\infty)={{1}\over {2}}k_c(\Gamma\rightarrow
\infty)$, is $\gamma_{max}(\Gamma\rightarrow\infty)=\pi G\sigma_0/a$. (Again,
in these dimensional expressions, $a$ is dimensional).

The maximum growth rate, and the wavenumber at which it occurs, is always less
for a magnetized but supercritical cloud than for an unmagnetized cloud, and,
as expected, the supercritical case approaches the unmagnetized case as the
magnetic fieldstrength decreases to zero.

In this paper we are interested in clouds which are magnetically subcritical,
so that they would be stable in the limit $\Gamma=0$, but would be unstable
if the magnetic field were removed.  That is, we are
interested in clouds with a length much
larger than the unmagnetized Jeans length. A small but nonzero $\Gamma$ 
destabilizes a cloud to
perturbations which are stabilized by magnetic fields in the absence of
ambipolar drift, but would be unstable to the Jeans instability in the
absence of magnetic fields.

%%%
For small sound speeds, the dispersion relation shows the same 
behavior seen for zero temperature in Z98: 
at low values of the drift parameter 
$\Gamma$, the growth rate of the perturbation $\gamma$ is 
proportional to $\Gamma$. At higher $\Gamma$, however, 
$\gamma\propto\Gamma^{1/3}$. As surface densities $\omega$ and wavenumbers
$k$ depart from the critical values for stability, more ambipolar drift is
required for the system to show $\gamma\propto\Gamma^{1/3}$ behavior. 
To quantify these
statements with an example, at $\omega_0$ = 1.1, $a^2$ = 0.02, $\Gamma$ = 0,
the critical $k$ below which ideal perturbations are unstable is $k$ = 9.5455.
Very close to criticality, $k$ = 9.6, the $\gamma\propto \Gamma^{1/3}$
scaling holds for $\Gamma$ as small as .001. At $k$ = 10, $\gamma$ increases
with $\Gamma$ faster than $\Gamma^{1/3}$, but much slower than linearly, for $
.001 < \Gamma < .01$, but approaches the $\Gamma^{1/3}$ scaling for $.01 <
\Gamma < .1$. As $\Gamma$ is increased from .001 to .1, $\gamma$ increases
from .378 to 1.75, which is $\Gamma^{.33}$ scaling. At $k$ = 4$\pi$, this
scaling has broken down noticeably: as
 $\Gamma$ is increased from .001 to .1, $\gamma$ increases
from .170 to 1.93, which is $\Gamma^{.53}$ scaling. Most of the deviation
occurs for small values of $\Gamma$; for $\Gamma$ between .01 and .1, $\gamma
\propto \Gamma^{.36}$. Thus, for reasonable values of $a^2$ and $\Gamma$, the
$\Gamma^{1/3}$ scaling law holds quite well even when $k$ is
as much as 30\% below the
critical value.

\begin{figure}[hbpt]
\centerline{\resizebox{\colw}{!}{\includegraphics{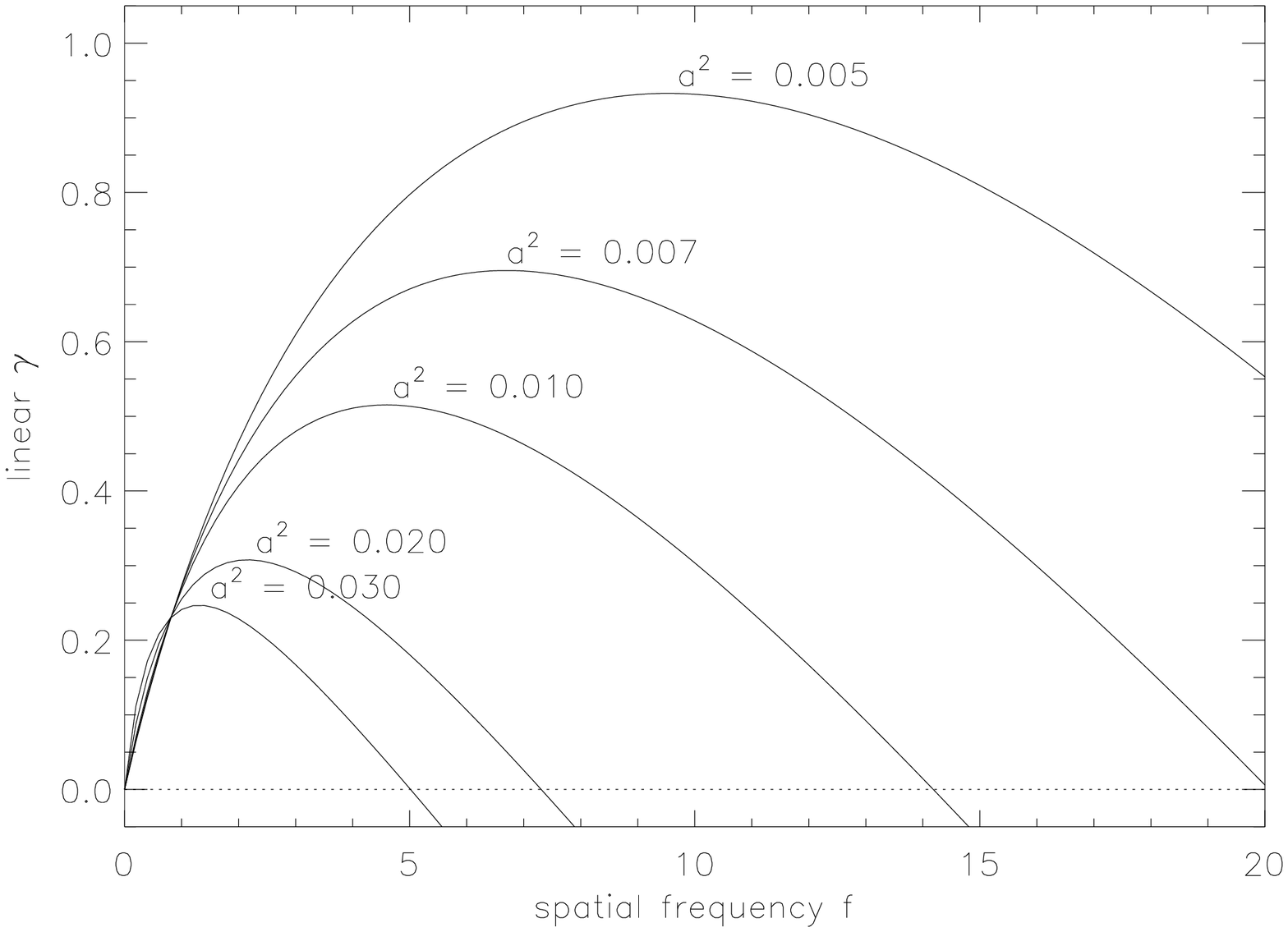}}}
\centerline{\resizebox{\colw}{!}{\includegraphics{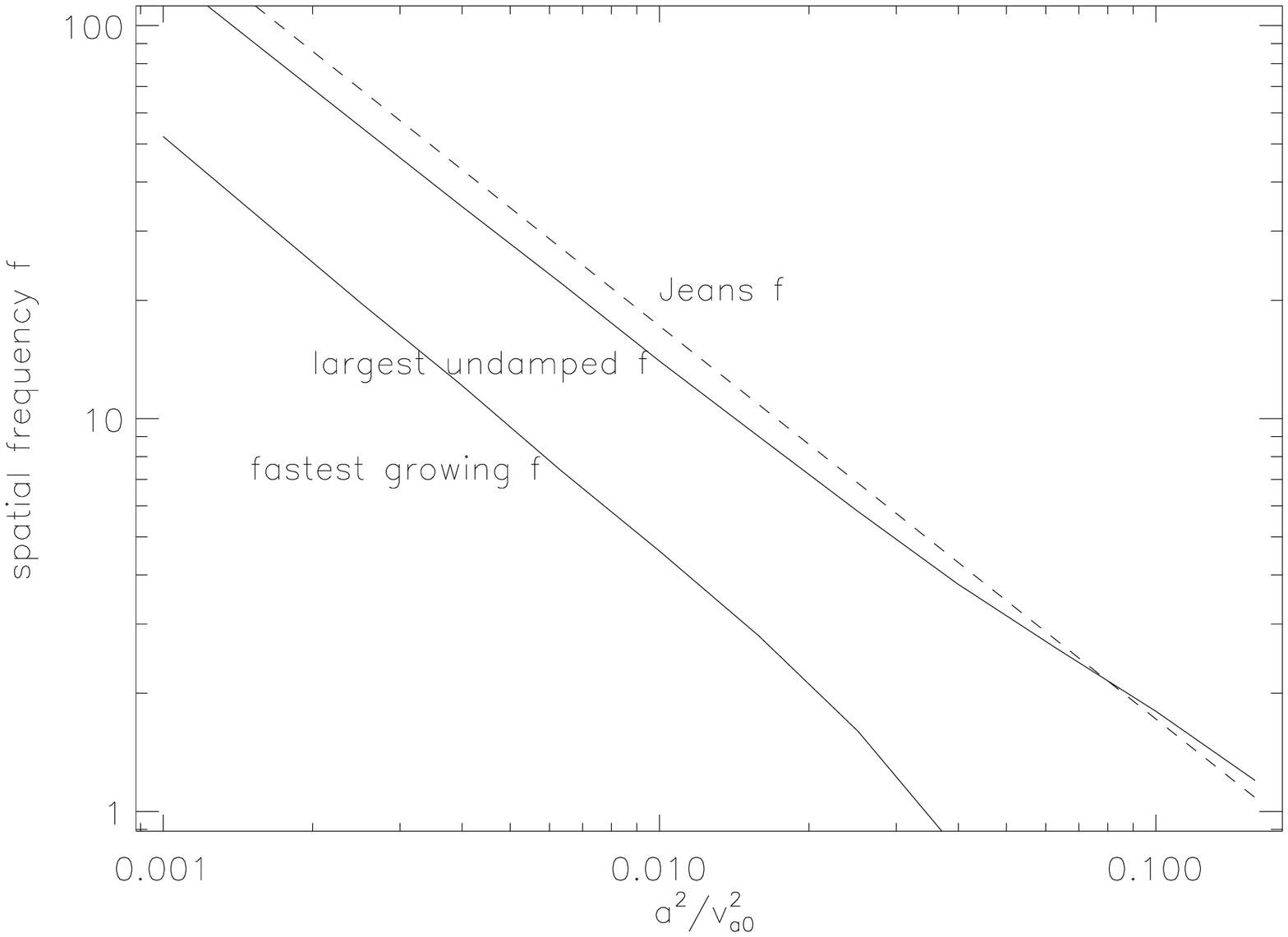}}}
\caption{\label{ak}
Damping of short-wavelength modes due to thermal pressure. The 
first graph shows the linear growth rate $\gamma$ of the fastest
growing mode as a function 
of spatial frequency $f=k/2\pi$, for different sound speeds $a^2$. 
The second graph shows the spatial frequency of maximal growth 
rate (the peak of each curve in the first graph), and the highest 
undamped spatial frequency (where each curve in the 
first graph crosses the x-axis). 
For both figures, $\Gamma$ = 0.1, and $\omega_{crit}$ is approximately 
constant: $\omega_0$ = 0.864(1+$\pi a^2$).
}\end{figure}

The addition of thermal pressure increases the stability of the disk;
%and suppresses modes with high spatial wavenumbers. 
more drift (larger $\Gamma$) is required for collapse, and more 
is required to reach the 
transition from $\gamma\propto\Gamma$ to $\gamma\propto\Gamma^{1/3}$. 
An approximate value for the critical surface density 
for collapse, with positive 
thermal pressure, is obtained from solving the dispersion relation 
(see eq. [\ref{disp_rel}]) for $\Gamma$=0:
\begin{mathletters}\begin{eqnarray}
{{1}\over{\omega_{\rm crit}}} 
	&=& -\pi a^2 + \sqrt{1+\pi^2a^4} \\
	&\simeq& 1-\pi a^2, \nonumber\\
\omega_{\rm crit} &\simeq& 1 + \pi a^2,
\end{eqnarray}\end{mathletters}
where the approximations hold for small sound speed.
Solution of the dispersion relation also shows that for a given sound 
speed and drift parameter, there is a single mode with a
maximal growth rate, and that the modes above a certain wavenumber
are acoustically suppressed.  Figure \ref{ak} shows the 
growth rate as a function of wavenumber for $\Gamma$ = 0.1, and most
unstable wavenumber, over a range of sound speeds. 
Figure \ref{ak} also shows that the stability boundary is very near the 
thermal Jeans stability boundary, while the fastest growing mode has a
much longer wavelength than the Jeans wavelength.
%%%

As we will see later, the wavenumber at which the growth rate is maximized
dominates the structure of clumps even into the nonlinear regime. Numerical
solution of the dispersion relation equation (\ref{disp_rel}) shows that $k_m$
is always less than $k_m(\Gamma\rightarrow\infty)$, the fastest
growing wavenumber for the Jeans instability, but also scales with temperature
as $1/T$. Therefore, we expect the fragment mass to be larger than the
thermal Jeans mass, but to have the same $T^2$ temperature scaling. This is
borne out by Figures \ref{ak} and \ref{clumpfig}.

%%%%%%%%%%%%%%%%%%%%%%%%%%%%%%%%%%%%%%%%%%%%%%%%%%%%%%%%%%%%%%%%%%%%%%%%
\section{Numerical Simulation}\label{num}

In order to follow the instability into the nonlinear regime we have 
carried out a numerical simulation.
We use a Fourier collocation pseudo-spectral method \citep{canuto} to
solve the governing equations.  Values of physical quantities are stored 
at discrete points in physical space (known as collocation points), and 
spatial derivatives are evaluated in Fourier spectral space (hence the name
``Fourier pseudo-spectral'').  
This particular method is well adapted to this
problem for several reasons.  Calculating a spatial derivative in Fourier 
space simply requires multiplication of each Fourier component by $i$ times
its wavenumber.  All terms involving the horizontal magnetic field 
$\mathbf{B}_h$ or gravitational potential $\Phi_G$ are trivial to evaluate
in Fourier space due to the simple form of the magnetic and gravitational
potentials.  Nonlinear terms, on the other hand, are trivial to evaluate in
physical space by simple multiplication.  Finally, growth of different 
Fourier modes can be monitored and controlled explicitly, simplifying 
comparison between the nonlinear numerical model and the single-wavenumber
linear analytic results.

We use a Bulirsh-Stoer time-stepping routine with Richardson extrapolation
\citep{fourn2}.  The routine performs several modified midpoint method
integrations at sub-intervals of the desired time-step.  It then attempts 
to extrapolate to an infinite number of sub-intervals.  The routine varies
the number of explicit (calculated) sub-intervals based on the estimated 
error.  In general, the full set of governing equations for this problem
can be integrated with $\sim$ 5 explicit subintervals for each time-step of
$\delta\tau = 0.1$.

\begin{figure}[hbt]
\centerline{\resizebox{\colw}{!}{\includegraphics{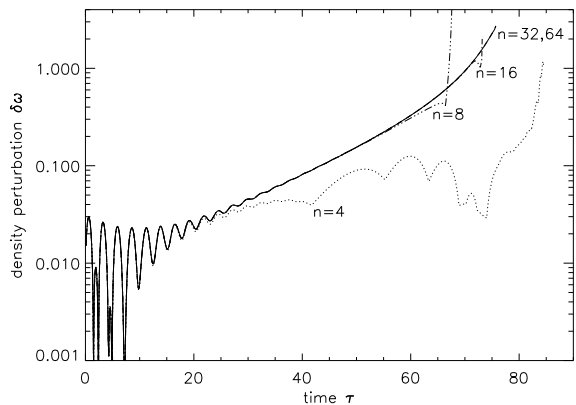}}}
\caption{
\label{conv}
Convergence of model with increasing numerical resolution.
Magnitude of the density perturbation for the same initial 
conditions, using grid sizes of $4^2$, $8^2$, $16^2$, $32^2$, 
and $64^2$.
This particular run had $\Gamma$ = 0.1, $a^2$ = 0.1, 
$\omega_0$ = 0.864 $\omega_{crit}$ = 1.135). The oscillations at early times
are due to the presence of frictionally damped oscillatory modes which were
present in the initial conditions.}
\end{figure}

We tested  convergence by running the code at increasing 
spatial resolution 
with the same initial conditions. Figure \ref{conv} shows 
the amplitude of a density perturbation, computed at
different resolutions, as a function of time. 
The initial conditions had a single wavenumber density perturbation
in each direction, 
forming a ``checkerboard'' pattern. (Collapse is described in 
detail below.)
A 16$^2$ grid is sufficient to resolve the collapse of 
such a single wavenumber, from the linear regime (exponential
growth) into the nonlinear regime.
Use of a 32$^2$ or 64$^2$ grid changes the solution by less than 0.1\%
over most of the time period plotted.  Finer grids are required to 
resolve collapse of smaller structures, and in runs which 
contain a spectrum of wavenumbers, a 64$^2$ grid was used.

\begin{figure}
\centerline{\resizebox{\colw}{!}{\includegraphics{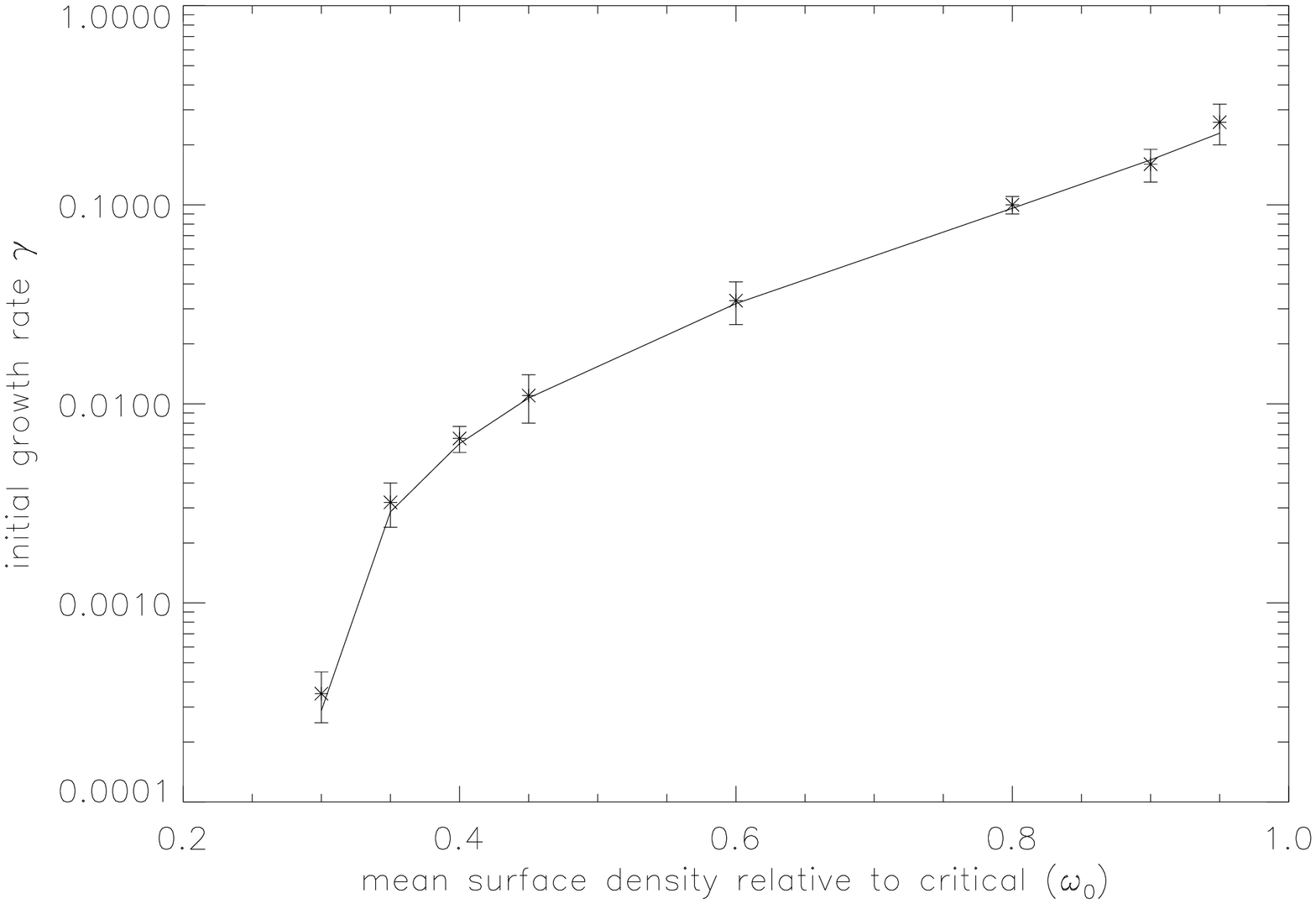}}}
\centerline{\resizebox{\colw}{!}{\includegraphics{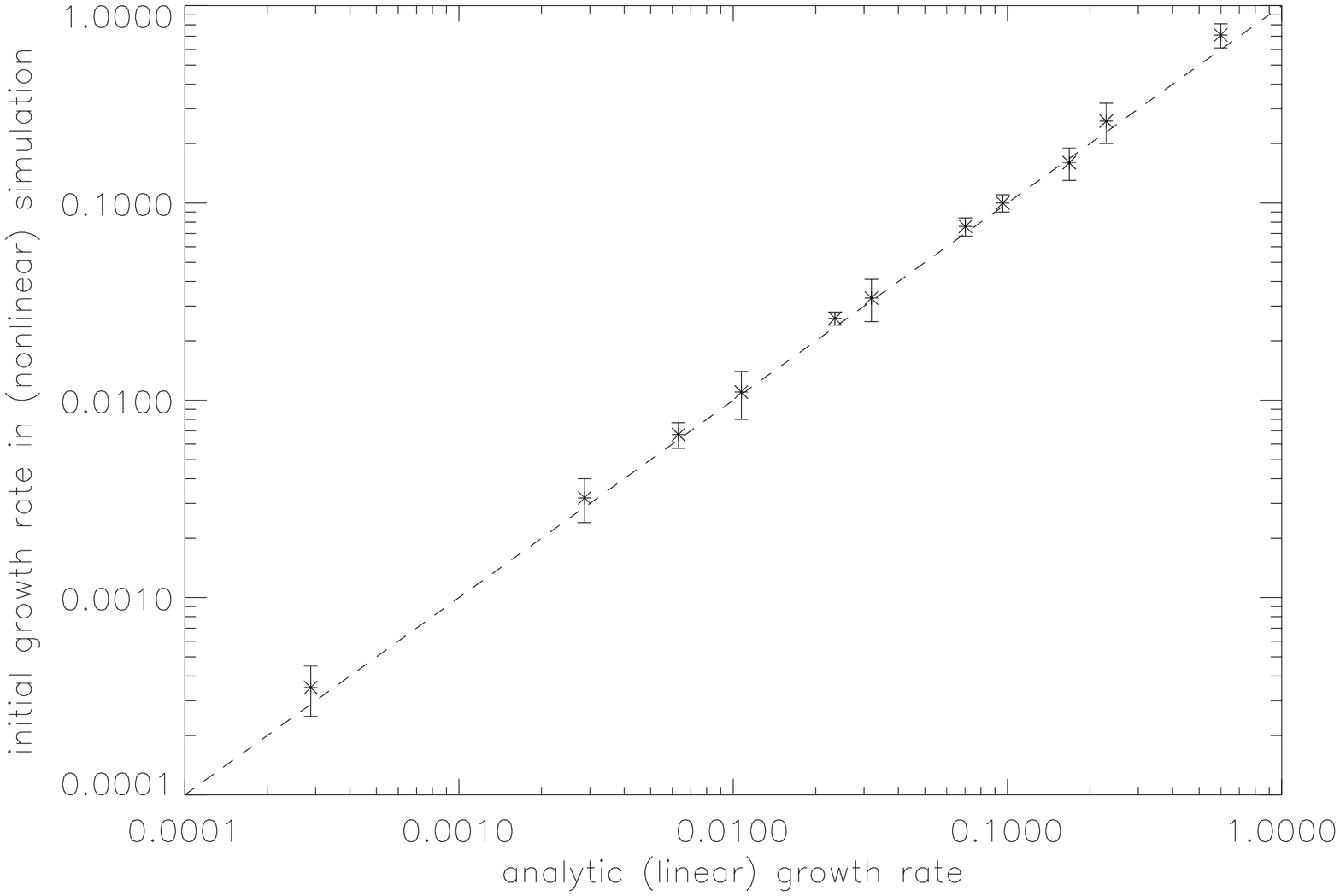}}}
\caption{
Comparison of linear analytical growth rate (solid line) with the 
initial growth rate in our simulations 
($\gamma$ when the density perturbation $\delta\omega$ is 
$<$10\% of mean density $\omega$), for a 
representative sound speed $a^2$ = 0.033.
Left, the growth rate in the simulation (points with error bars)
agrees with the linear theory (solid line) for 
a range of initial surface density $\omega_0$ ($\Gamma$ = 0.1). Right,
 the growth rate in the simulation is seen to agree 
with the linear theory over a 4 order of magnitude range in $\gamma$.
($\Gamma$ varies from 0.01 to 0.3 and $\omega_0$ from 0.3 to 0.95)
\label{lincomp}}
\end{figure}

We have also verified that the code reproduces the results of linear theory.
If the initial condition corresponds to an eigenfunction of a growing mode
calculated according to linear perturbation theory, with an amplitude of a few
percent or less, then the disturbance initially grows at the exponential rate
predicted by the linear theory. This is shown in Figure \ref{lincomp}, which
compares the growth rates measured from the code (discrete symbols) with the
continuous curve obtained from solving the dispersion relation 
equation (\ref{disp_rel}).
The agreement is generally excellent. 

The model is numerically stable until such time as power in the higher 
order wavenumbers grows to overwhelm power in the Fourier modes of 
interest. 
At that time, the simulation develops a ``sawtooth'' instability, 
with large variation between alternating collocation points. 
Power grows in these short-wavelength modes from numerical 
noise, whose magnitude is about 10\textsuperscript{-8} compared to the 
power in the principal mode (measured in simulations whose initial 
conditions contained a single mode).  Higher wavenumber modes are
also driven by the nonlinearity of the problem, and this is the
dominant physical source of power in those modes.  

Growth of high-order modes can be controlled in several ways.  
Thermal pressure will stabilize high order modes, as was seen 
in Figure \ref{ak}.  In most cases, a physically reasonable 
finite cloud temperature of $\lesssim$ 10 K will stabilize the 
simulation long enough to follow the collapse well into the nonlinear regime.
The problem of high-order mode stabilization 
is nearly independent of the drift parameter $\Gamma$ because
an increase (decrease) in ambipolar diffusion increases
(decreases) the growth rate of all modes similarly. 
Thus the entire physically interesting part of parameter space is
numerically accessible and numerically stable in this model.

%%%%%%%%%%%%%%%%%%%%%%%%%%%%%%%%%%%%%%%%%%%%%%%%%%%%%%%%%%%%%%%%%%%%%%%%%%%
\subsection{Collapse Rate}\label{gamgam}

Many runs were computed
with initial conditions corresponding to the eigenfunction 
of the fastest-growing solution of the 3 modes present, at each
wavenumber, in linear theory.  This initial perturbation has a  
single initial wavenumber in each direction.  
Growth initially proceeds exponentially, with faster
nonlinear collapse occurring as higher order spatial modes 
are driven.  Nonlinear behavior is typically seen when the 
density enhancement associated with a perturbation has 
grown to 75\%-100\% of the mean density. We were able to follow the evolution
of the system to peak surface densities about 10 times larger than the mean
density (corresponding to a peak volume density about 100 times larger than
the mean).

\begin{figure}[hptb]
\centerline{\resizebox{\colw}{!}{\includegraphics{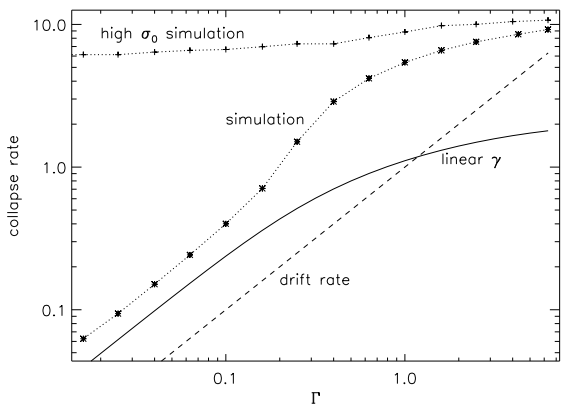}}}
\caption{Collapse rate for single-wavenumber initial conditions, 
$a^2 = 0.033$, and $\omega_0 = 0.954 = 0.864$ $\omega_{\rm crit}$
(lower three curves).
These initial conditions would be stable in the absence of 
ambipolar drift.
The solid line is the linear growth rate $\gamma$ as a function of 
the strength of the ambipolar drift $\Gamma$.  
The stars connected by a dotted line is the collapse rate for the nonlinear
simulation.  The dashed line is the ambipolar drift rate. 
The pluses connected by a dotted line 
are the collapse rates for the nonlinear 
simulation with a high initial surface density
$\omega_0 = 1.278 = 1/0.864$ $\omega_{\rm crit}$.  For such supercritical 
collapse, the magnetic field and strength of ambipolar drift should have 
minimal effects.
\label{tau}
}\end{figure}
We made a detailed comparison with linear theory and 
with simple drift and collapse models by
studying the early collapse of a perturbation with a 
single spatial wavenumber.
Figure \ref{tau} shows how the growth of a fully nonlinear 
density perturbation depends on the drift parameter
$\Gamma$.  The lower three curves describe collapse in 
subcritical clouds, which would be stabilized by the 
magnetic field in the absence of ambipolar drift.  
The initial surface density (normalized to the vertical magnetic field) 
is $\sim$ 86\% of the critical surface density for collapse.  
The growth rate of perturbations in the nonlinear simulation 
(calculated from the time for the central density of the 
perturbation to grow from 1\% to 100\% of the mean density) is 
compared to the predicted growth rate $\gamma$ for linear perturbations, 
and to the ambipolar drift rate $\Gamma$.  
Clearly, for the physically expected value
of $\Gamma$ ($\sim$ 0.05 - 0.10), \S\ref{phys_par}), the collapse due to 
this instability is several times faster than simple collapse due to 
loss of magnetic support on a diffusive ambipolar drift time-scale.
For
example, when $\Gamma$ = 0.1, the ambipolar drift rate is 0.1 and the
unmagnetized collapse rate is 10. The rate of contraction found from the
simulation is 0.4. At larger values of $\Gamma$, the drift rate comes closer to
the unmagnetized collapse rate. The collapse rate in the simulation approaches
the unmagnetized collapse rate because the coupling between the magnetic
field and the gas is weak. 
When the drift becomes very important ($\Gamma\sim 10$), the drift 
timescale is very short, and the intermediate instability described in this
paper is 
less significant. Even for the moderate sized density perturbations used to
create Figure \ref{tau}, ($\delta\omega\sim\omega$), the growth rate is  
larger in the nonlinear simulation than in the
linear problem, especially at relatively large values of
$\Gamma$, showing that the collapse is accelerated by the 
nonlinearity.
It is important to note, however, that the growth rate shows the
same dependence on ambipolar drift in both the linear theory
and the nonlinear simulation: at low $\Gamma$,
$\gamma\propto\Gamma$, and at higher $\Gamma$,
$\gamma\propto\Gamma^{1/3}$.

The top curve in Figure \ref{tau} describes collapse in supercritical
clouds, in which the magnetic field would be insufficient to prevent
collapse even if it were frozen to the matter.  The collapse
rate depends only weakly on the strength of the ambipolar
drift, as expected since the magnetic field is dynamically less
important.  When the ambipolar drift strength $\Gamma$ becomes
large, the drift timescale becomes comparable to the
collapse timescale, and the subcritical and supercritical
cases converge.  Rapid collapse occurs in supercritical
clouds due to the dynamical weakness of the field, and
in subcritical clouds rapid diffusion removes magnetic
support, quickly rendering them supercritical.

The evolution of self gravitating, subcritical disks
with ambipolar drift was studied previously by \citet{cm94}
and \citet{bm95}. \citet{cm94} began with a 
subcritical ($\omega_0$ = .256),
centrally condensed equilibrium state - the central surface density is 16
times the mean density. Thus, this model is more centrally condensed even
initially than our models are when we terminate the simulation. 
The initial ambipolar drift time is 10 times the
initial free fall time, which corresponds on Figure \ref{tau} to
$\Gamma\sim 1$. The evolutionary timescale in the subcritical, 
quasistatic phase is well estimated by the initial ambipolar drift time; after
the cloud becomes supercritical its collapse rate approaches the freefall rate.
Although we can extrapolate our results to this model only with caution,
because the initial conditions are so different from ours, it does not
surprise us that such a subcritical disk shows no evidence for the blending
of dynamical and drift effects that we observe closer to criticality.

\citet{bm95} carried out a parameter study to determine the effects of the
degree of criticality on the rate of evolution to a critical state. They
also began with centrally condensed equilibrium models, forming a sequence in
which the criticality parameter varied from 0.1 to 0.5. They found that the
timescale for evolution to the critical state decreased by about a factor of
1.5 along this sequence, from somewhat longer than the estimated drift time
to about 25\% shorter (another, marginally critical model, collapsed at once). 
Although again a quantitative comparison of our models
with theirs is difficult because of the different initial conditions, it
is possible that the
intermediate contraction rates which they see are a manifestation of the
coupling between dynamical and ambipolar drift effects seen in our models.
It may also be due to the increased central concentration of the initial
equilibrium states along their sequence of models.

The nature and rate of collapse is observable in molecular clouds
\citep{evans99,meo99}, 
and an instability with an intermediate growth rate 
such as this one can help to explain observations that do not 
fit either of the classical scenarios - dynamical collapse 
or slow diffusive contraction.
Our simulated cloud cores collapse 
with slower velocities and on larger physical scales 
than the dynamical inside-out collapse predicted when the 
magnetic field is unimportant, as in \citet{shu77}. \citet{t98} and 
\citet{greg} have observed cores that appear to have such behavior; they find 
that the regions of inflow are too large to fit dynamical 
inside-out collapse models, but that the inflow velocities 
are too large for quasistatic diffusion models.

%%%%%%%%%%%%%%%%%%%%%%%%%%%%%%%%%%%%%%%%%%%%%%%%%%%%%%%%%%%%%%%%%%%%%%
\subsection{{\bf B}-$\rho$ relation}\label{brho}

The correlation between fieldstrength and density is an observable quantity
which can provide insight into the manner in which the magnetic field 
evolves. It is
useful to parameterize this relation as a power law
\begin{equation}
|\mathbf{B}|\propto\rho^\kappa.
\end{equation}
Observations find $\kappa\simeq$ 0.5 \citep{troland86,c99} over
several orders of magnitude in density.
In the case of a highly flattened cloud such as we simulate here, 
it is more convenient to define a
magnetic field - surface density relation
\begin{equation}
 |\mathbf{B}|\propto\sigma^\lambda. 
\end{equation}
If the slab scale height $H$ remains constant throughout the 
evolution, then $\lambda=\kappa$.  If the scale height is determined 
by a balance between self-gravity and thermal pressure alone, 
then an isothermal slab obeys $H\sim 1/\sigma$, and 
$\sigma\sim\rho H$ implies $\sigma\sim\rho^{1/2}$, or 
$\kappa=0.5\lambda$  \citep{c99,spit42}.

If the magnetic field is frozen to the matter but not dynamically 
important, so contraction is isotropic,
 conservation of flux $\Phi_{mag}\propto L^2|\mathbf{B}|$
and mass $M\propto L^3\rho$ requires $|\mathbf{B}|\propto\rho^{2/3}$
($\kappa=$\ 2/3). If the field is so strong that matter moves one
dimensionally, parallel to the fieldlines, then $\kappa\rightarrow 0$.
Calculations in which a cloud condenses to magnetohydrostatic equilibrium
from a uniform initial state, with frozen in magnetic flux of a magnitude
appropriate to the ISM, show anisotropic contraction, and the central values
of $B$ and $\rho$ in the
initial and final states are related by $\kappa\sim 0.5$ \citep{ms76,tin88}.
If a cloud is already flattened and shrinks transversely, $M
\propto L^2\sigma$, and flux
freezing implies $|\mathbf{B}|\propto\sigma^1$ ($\lambda$=1, $\kappa$=0.5). 
However, rather
different input physics leads to a similar exponent: simulations of supersonic
magnetized turbulence, without self-gravity, produce $\kappa\sim 0.4$ if
the field is not too strong \citep{pn99}. 

\begin{figure}[hbpt]
\centerline{\resizebox{\colw}{!}{\includegraphics{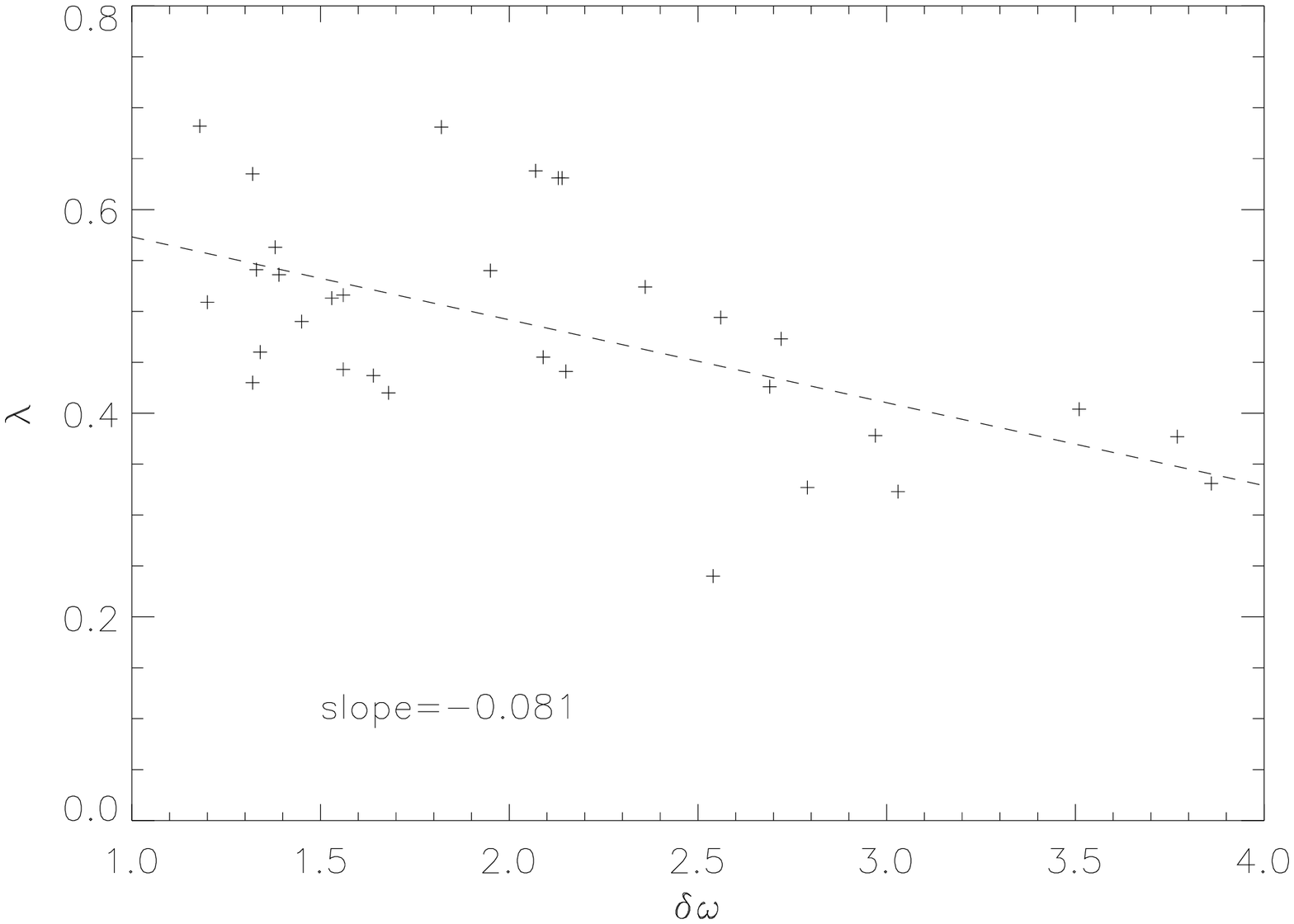}}}
\centerline{\resizebox{\colw}{!}{\includegraphics{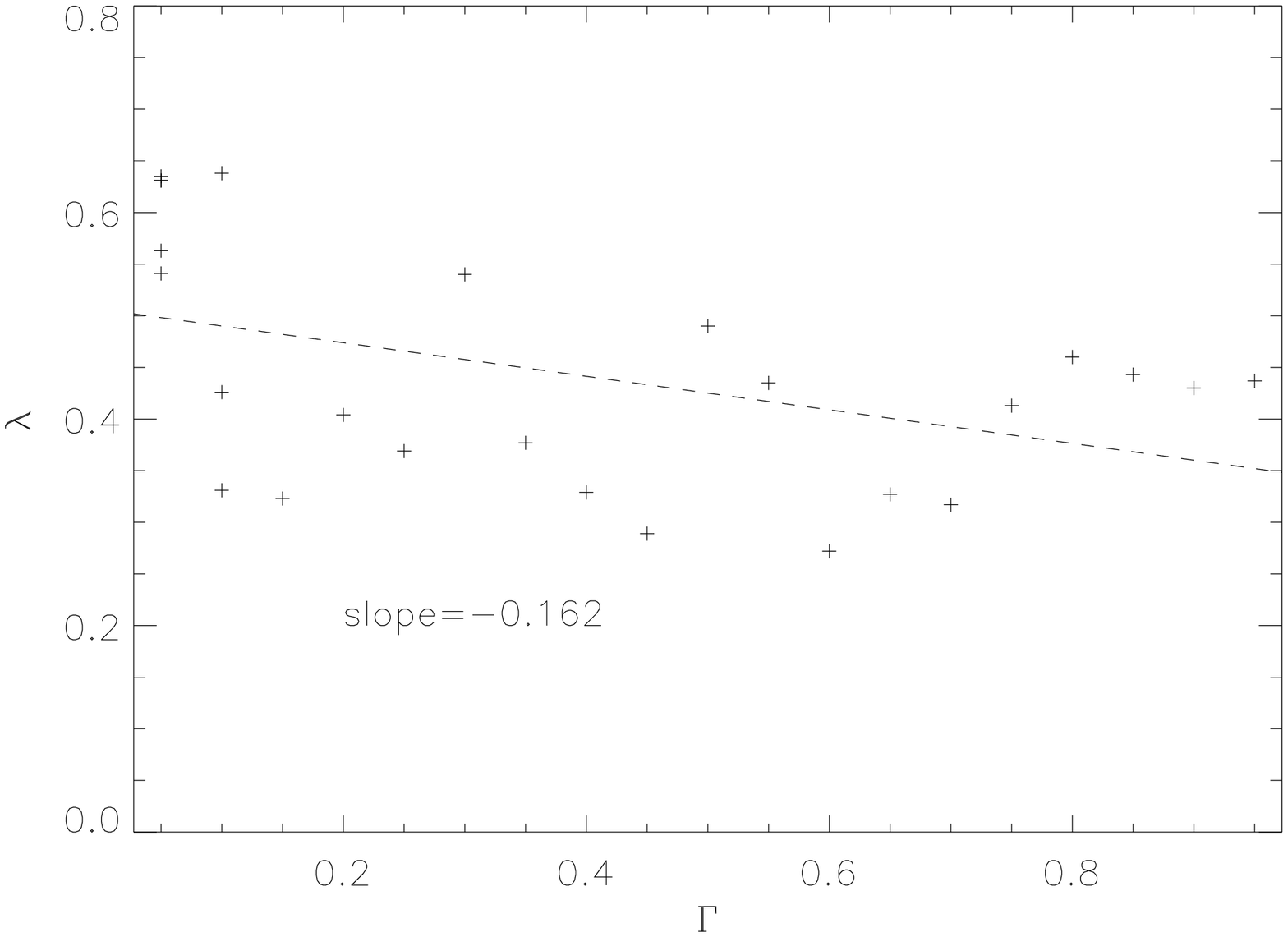}}}
\centerline{\resizebox{\colw}{!}{\includegraphics{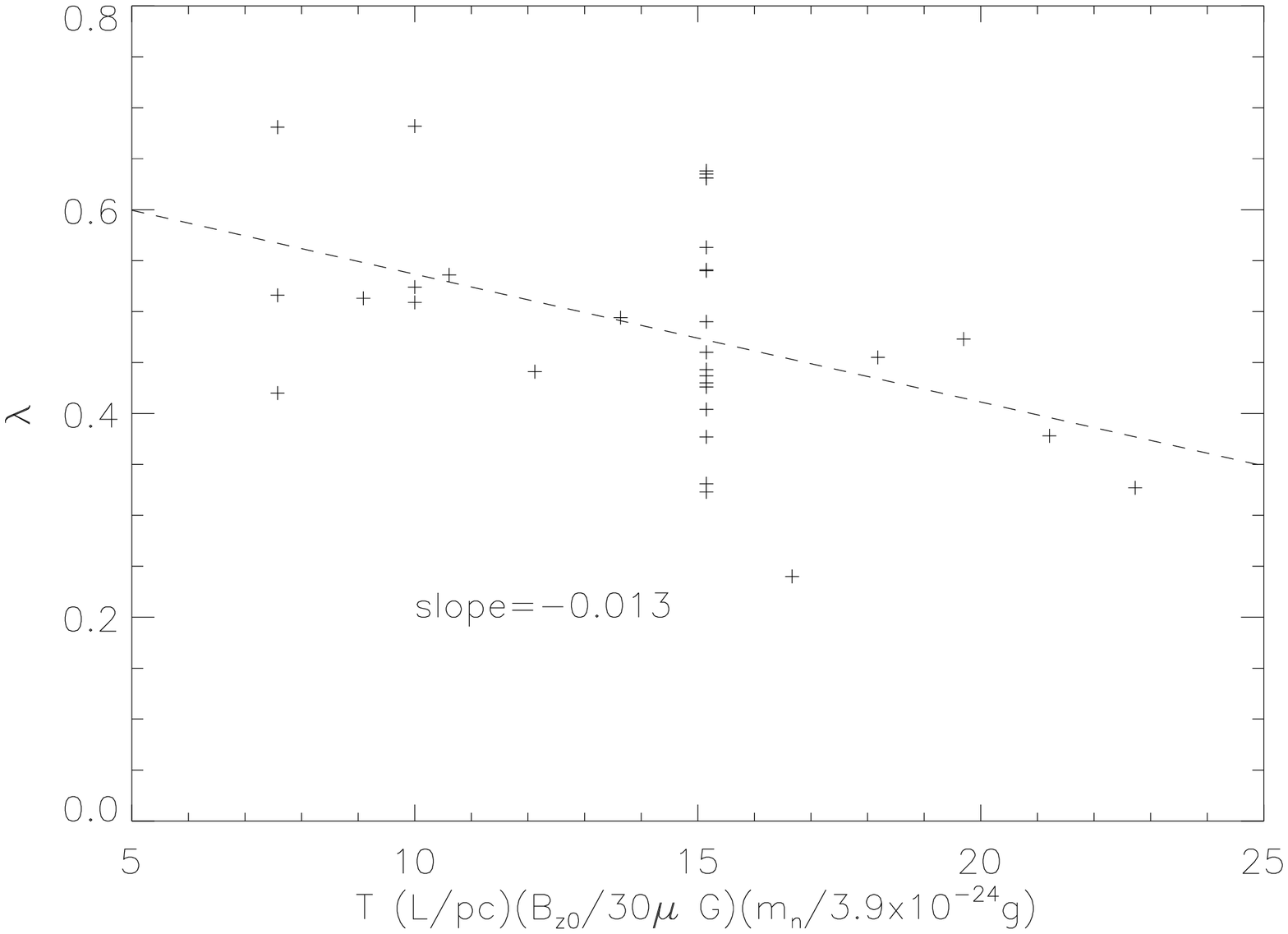}}}
\caption{
The magnetic field - density exponent $\lambda$.
($|\mathbf{B}|\propto\sigma^\lambda$).  The exponent 
decreases (increased flux loss from a collapsing clump)
with increasing central density $\delta\omega$, 
increasing ambipolar drift strength $\Gamma$, or 
increasing cloud temperature T.
The first graph shows variation with 
$\delta\omega$, the second variation 
with $\Gamma$ at constant $a^2=$ 0.05, and the third,
variation with T at constant $\Gamma=$ 0.1.
In all simulations, $\omega_0$ = 0.864 $\omega_{crit}$.
\label{brhofig}
}\end{figure}

Ambipolar drift generally reduces $\kappa$ below the value it would have if
the field were frozen in. In the models of \citet{fm93}, 
$\kappa$ averages
about 0.2 during the so-called quasistatic phase. After the quasistatic phase
ends, the mean value of $\kappa$ is 0.3 as $\rho$ increases by more than 5
orders of magnitude. In the highly flattened models of 
\citet{cm94}, $\kappa$
increases smoothly as the cloud evolves from subcritical to supercritical,
reaching peak values between 0.4 and 0.5 and being about 0.3 at criticality.
In our simulations the magnetic field - surface density exponent $\lambda$ 
varies between 0.35 and 0.65.  As shown in Figure \ref{brhofig}, 
the exponent decreases as the central density of a clump increases. 
As collapse proceeds, not only does the density increase, but the magnetic 
field curvature also increases as fieldlines are dragged into the 
condensation.  Both effects increase the ambipolar drift 
velocity $\mathbf{v}_D = B_z\mathbf{B}_h/2\pi\sigma\nu_{ni}$ 
and thus the rate of flux loss 
from the clump.  Our models do not show the increase of
$\lambda$ toward its frozen flux value as the central density increases
seen in \citet{cm94}, 
because we follow only the early stages of contraction, in which the
velocity is well below the freefall value.
The exponent $\lambda$ also decreases as the 
amount of ambipolar drift $\Gamma$ increases, as would be expected, and as the 
cloud temperature increases.  The latter effect results from 
the decreased efficiency of this instability in warm clouds. 
The collapse rate $\gamma$ decreases with increasing temperature
as shown in \S\ref{lin_th}, and the collapse time is longer 
relative to the ambipolar drift time, so more flux can leak 
from the clump as it collapses. 

Comparison with observations depends on the assumed disc 
scale height $H$ and the central densities of observed cloud cores.
\citet{c99} finds $\kappa$ = 0.47 for cloud cores with densities 
10\textsuperscript{2.5}cm\textsuperscript{-3} $\lesssim n_H\lesssim$
10\textsuperscript{7.5}cm\textsuperscript{-3}.  His data are 
strongly weighted by observations which only measure upper limits 
for the magnetic fieldstrength, and omission of those data points 
results in an exponent of 
$\kappa$ = 0.3 over the same range of densities. 
We measure 0.3 $\lesssim\lambda\lesssim$ 0.7 in simulated cores.
If the slab scale height $H$ is constant,  
then 0.3 $\lesssim\kappa\lesssim$ 0.7, but
if the slab obeys $H\sim 1/\sigma$, then 0.2 $\lesssim\kappa\lesssim$ 0.35.
Our simulations are thus
consistent with the small number of available observations.

%%%%%%%%%%%%%%%%%%%%%%%%%%%%%%%%%%%%%%%%%%%%%%%%%%%%%%%%%%%%%%%%%%%%%%
\subsection{Clumps and Fragments}\label{clump}
%%%%%
Real molecular 
clouds have density variations on many scales, or a spectrum of 
spatial wavenumbers.  The linear theory for this instability
(\S\ref{lin_th}) predicts that a single mode 
will dominate collapse.  Fragments of a single mass 
will form, with that mass depending on the temperature and degree of 
criticality, with little dependence on the 
strength of ambipolar drift, provided that it is small.  We have simulated 
many ($\sim$ 100) clouds in which the initial density perturbations 
have a broad spatial Fourier spectrum and random phase.
The real part of the 
initial Fourier spectrum of the density perturbation 
is a Gaussian centered on $k$ = 0 (with the omission of the $k$ = 0 
mode itself).  The FWHM is 0.33 $k_{max}$, where $k_{max}$ is 
the highest spatial mode in the simulation, so a range of low-order modes
have similar initial strength, and there is significant initial power 
even in some acoustically damped (high $k$) modes. 
After some time, the clouds coalesce into a small number of fragments, each
of which
we define to be a region with $\sigma > \sigma_0$.
The size of the fragments is well-predicted by the linear theory.
For example, when $a^2$ = 0.02, the 
fastest linearly growing mode is $k/2\pi \simeq 2.5$, and in the 
simulation, the modes ($k_x,k_y$)$/2\pi$ = (1,3) and (3,1) have
much larger amplitudes than other spatial modes.

\begin{figure}[hbpt]
\centerline{\resizebox{\colw}{!}{\includegraphics{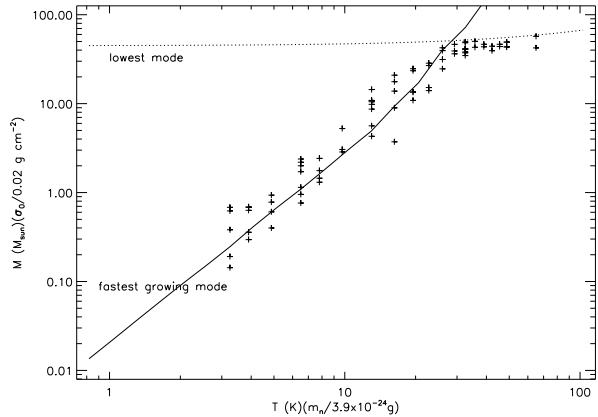}}}
\caption{\label{clumpfig}
Masses of cloud fragments as a function of temperature. 
Masses in the simulation (crosses) follow the mass of the 
fastest growing mode in linear theory (solid line) unless that 
mode exceeds the size of the simulation.  (see text)
These runs have $\Gamma$ = 0.1, $\omega_0$ = 0.864 $\omega_{crit}$.
Masses from the simulation are plotted assuming $L$ = 1 pc, but
the agreement between the mass predicted by linear theory and 
the mass in the simulation is independent of $L$.
}\end{figure}

Our simulations produce a wide variety of clumps and fragments, 
as expected for systems with random initial density fluctuations.
Figure \ref{clumpfig} shows the fragment masses at a fairly early 
stage of collapse (the density perturbation is $\sim$ 50\% greater
than the mean density).  There is considerable scatter, but the linear 
theory is a good guide to the average fragment mass.
The clump masses are similar to the masses 
of the supercritical cores formed in some previous simulations
\citep{fm93,bm94,cm94,ck98}.

At sufficiently large $T$, 
the spatial mode with the largest linear growth rate is the fundamental 
mode in the simulation domain 
($\lambda=L$), and fragments larger than this cannot form 
in a periodic simulation. 
This explains the apparent flattening of the clump mass --
temperature relation seen at the higher temperatures, as clump masses
are bounded above by the mass of that ``lowest mode''.

The question of whether there is a characteristic mass for molecular cloud
substructure is an important one. Several observational studies have found
a power law distribution of clump masses over several orders of magnitude, 
$dN/dM\propto M^{-p}$, where $p\sim
1.5 - 1.7$ \citep{b93}. \citet{k98} present evidence that the power law extends
far below 1 $M_{\odot}$.
However, in the Taurus molecular cloud there is evidence 
of a minimum scale of a few tenths of a parsec, corresponding to several solar
masses \citep{bw97}. \citet{g98} have argued for an inner scale of 0.1 pc,
which they identify with a transition to what they term ``velocity coherence".
These inner scales are of the same order as the thermal Jeans length, and
also close to the cutoff wavelength below which Alfv\'{e}n waves are critically
damped due to strong ambipolar drift \citep{mckee93}. 
Our results suggest that there is
another scale, which is somewhat larger, in magnetically subcritical clouds
with weak ambipolar drift.
%%%%%%%%%%%%%%%%%%%%%%%%%%%%%%%%%%%%%%%%%%%%%%%%%%%%%%%%%%%%%%%%%%%%%%%
\subsection{Velocity Structure}\label{vel}

Unlike axisymmetric collapse models, these simulations allow 
the study of asymmetric collapse, clumps with complicated 
morphology, relative motion of clumps, and their internal 
velocity and vorticity fields.  We find that collapse is often 
asymmetric and that significant vorticity is generated by the 
instability (although of course the net angular momentum remains zero
in our simulations; its absolute value is more than an order of magnitude 
smaller than the estimated numerical errors).

Figure \ref{swirl} shows four examples of collapsing fragments. The left
panels show contours of constant surface density, arrows indicating the local
direction and magnitude of velocity, and bold arrows indicating the center of
mass motion of each clump. The velocity field shows infall towards clumps,
but there is also a visual impression of ``swirling" or rotational motion.
This is borne out by the right panels of Figure \ref{swirl}, which show
contours of constant vertical vorticity overlaid on contours of constant
density. We show below that the magnetic field generates local vorticity.

The velocity field 
indicates that the collapse is in most cases asymmetric, with {\it e.g.} much 
greater infall velocities on one side of the clump than the other, and
significant nonradial motion. 
Clump mergers are possible - one is quite likely taking place in the bottom
panel of Figure \ref{swirl} - but the bulk motions are significantly slower
than the infall velocities internal to clumps. The reverse is generally true
in molecular clouds \citep[{\it e.g.} ][]{b93}, 
and this result in our simulations is
a manifestation of the fact that the velocity field in the system is weak.
We return to this point in \S 3.5. Sometimes the clumps move apart, but this
is ambiguous in a periodic domain. The infall velocities within individual
clumps are of order 0.25 $a$, and are more consistent with observations 
\citep{evans99,meo99}.
\onecolumn
\begin{figure}[p]
\centerline{\resizebox{\colw}{\colw}{\includegraphics{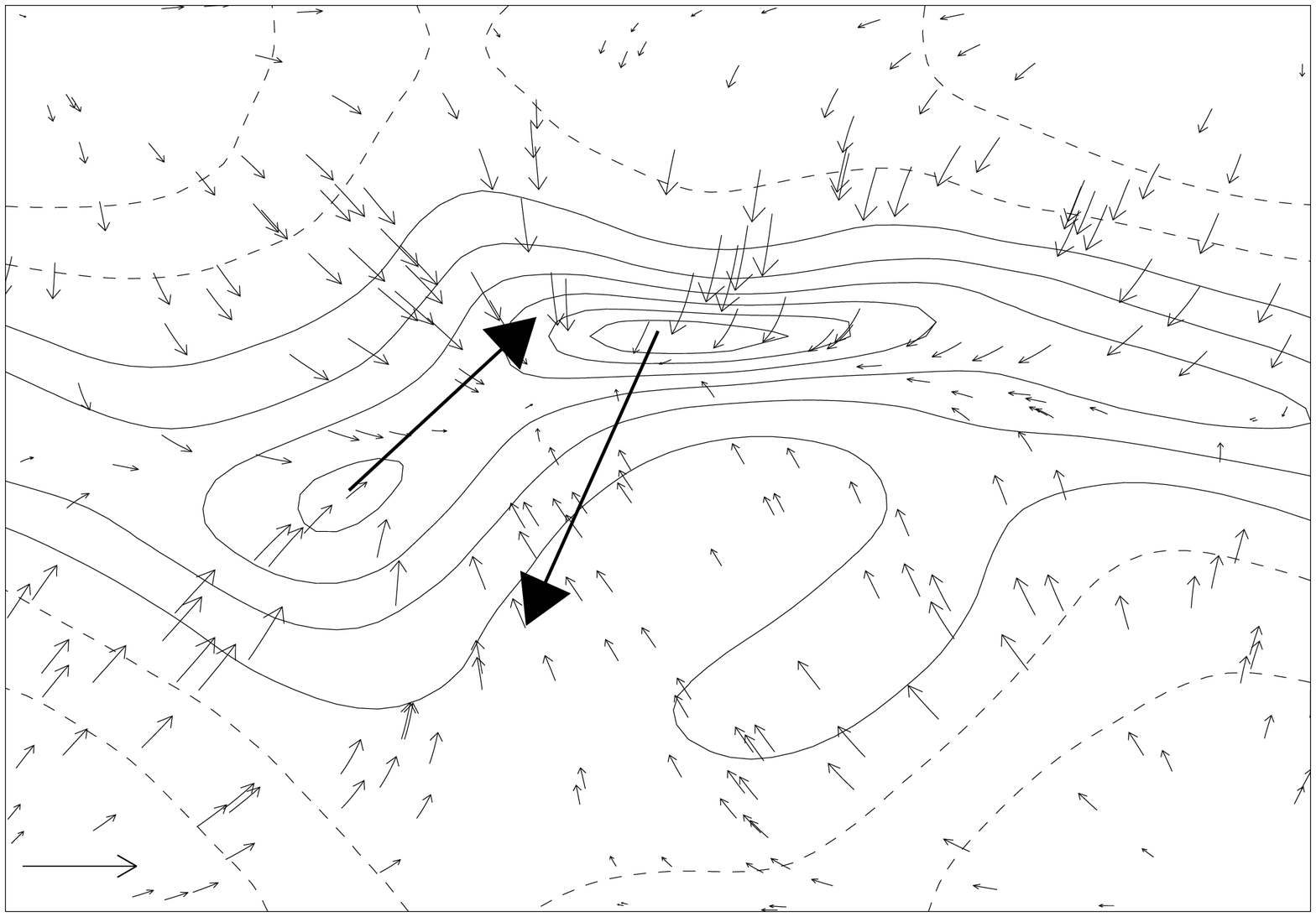}}
	\resizebox{\colw}{\colw}{\includegraphics{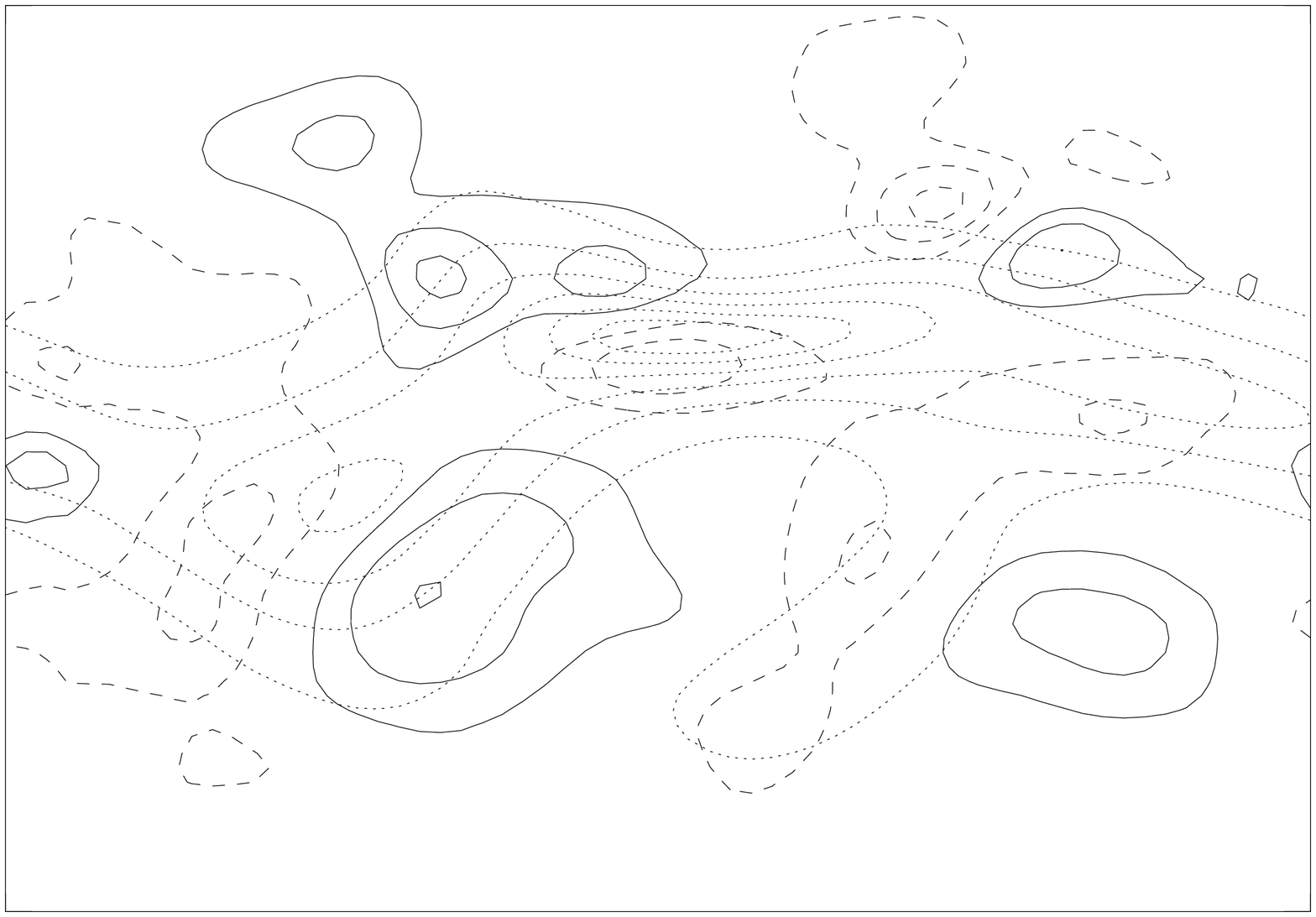}}}
\centerline{\resizebox{\colw}{\colw}{\includegraphics{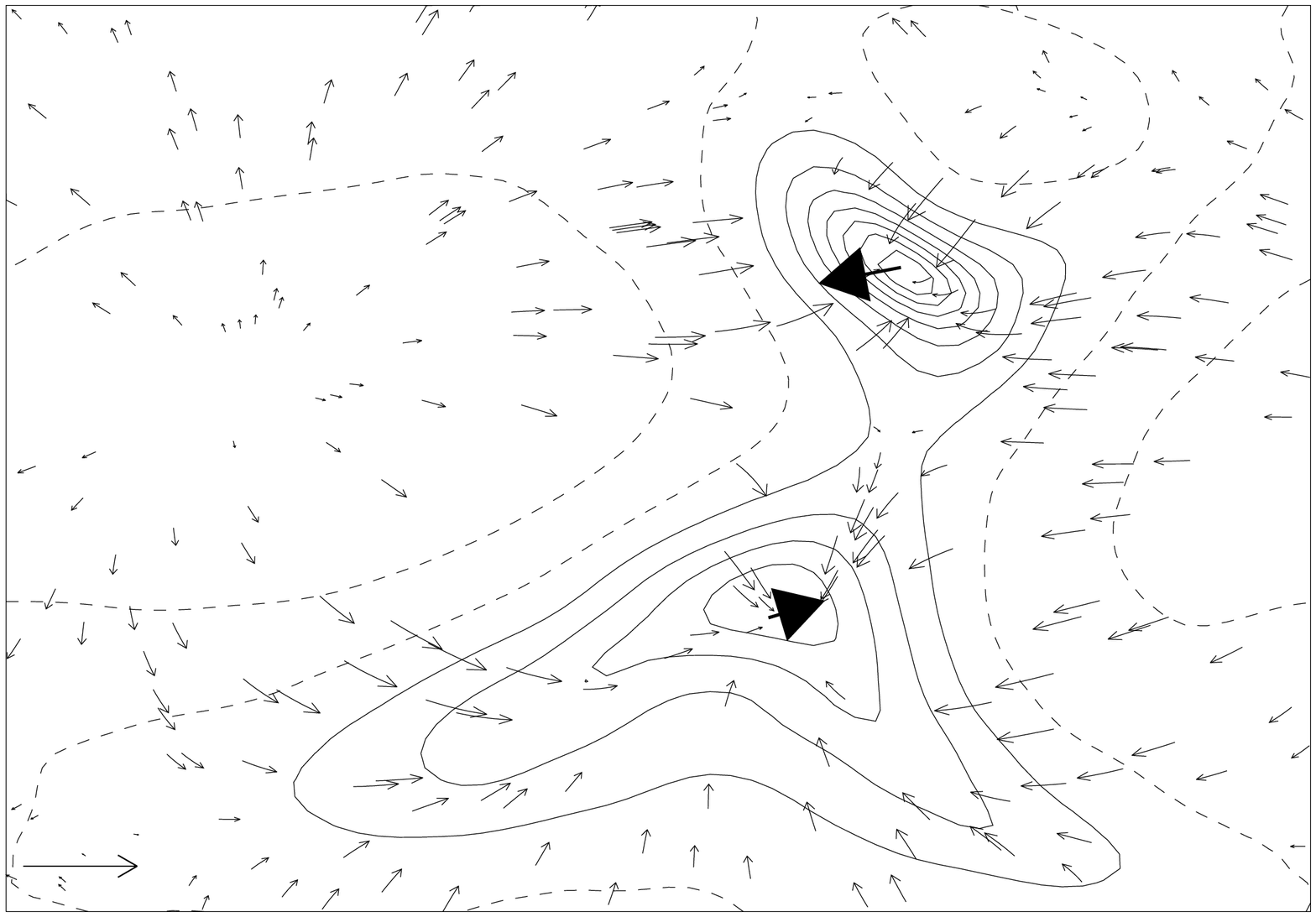}}
	\resizebox{\colw}{\colw}{\includegraphics{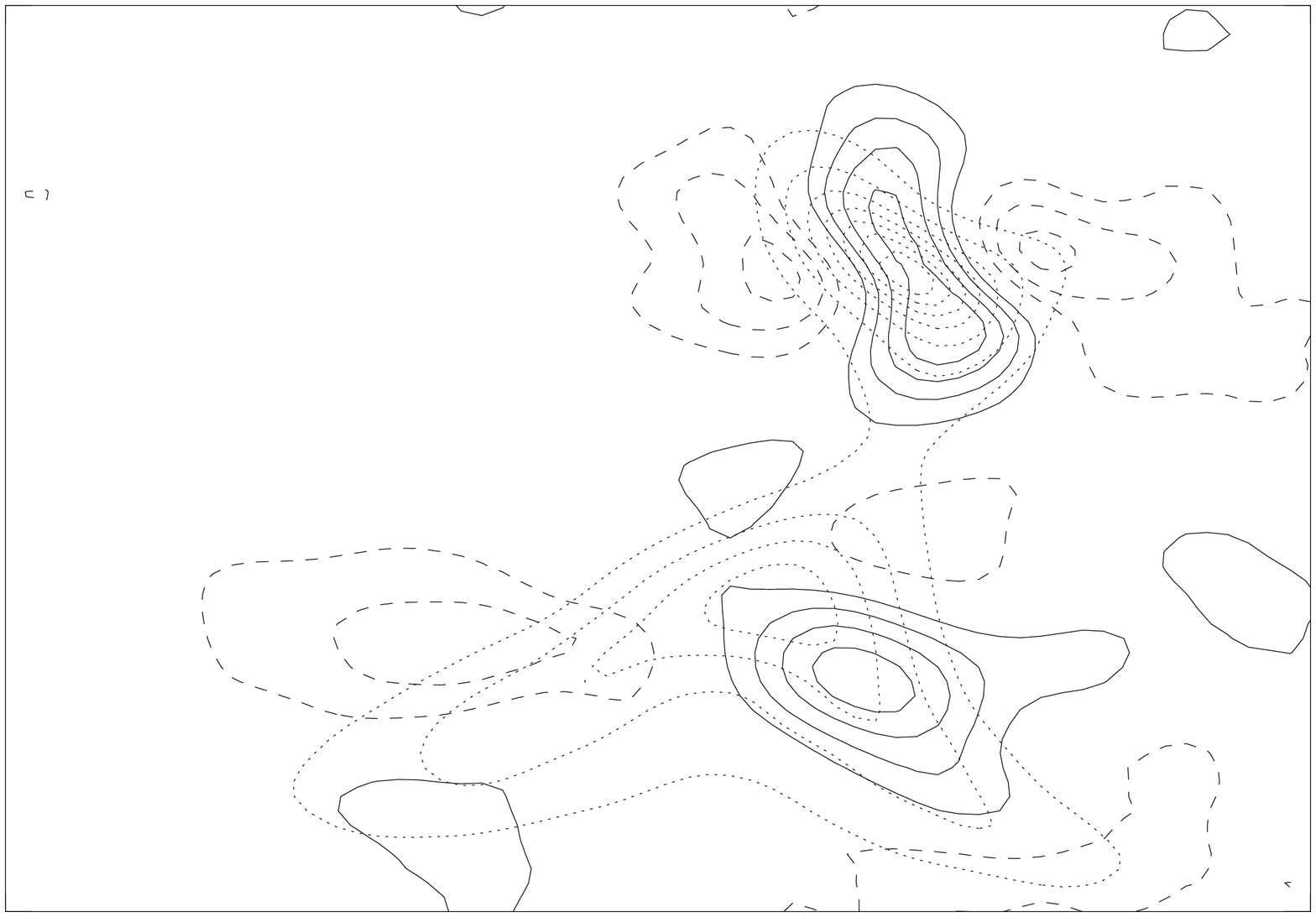}}}
\caption[Velocity and Vorticity]{
\renewcommand{\baselinestretch}{1}
\small\normalsize
Velocity, density, and vorticity for some typical simulations.

The left panel of each row shows the velocity field and density contours.
Contour levels denser than the mean density are solid, 
those less dense than the mean are dashed.
The velocity scale is indicated by the arrow in the lower left hand 
corner, whose length is 25 m/s.  The density-weighted mean velocity 
of each clump is indicated with a thick arrow, and the scale is 10 
times the general velocity scale (the arrow in the lower left would 
represent 2.5 m/s)

The right panel shows contours of vertical vorticity 
$(\mathbf{\nabla}\times\mathbf{v})_z$ for the same cloud.  Positive 
(solid) and negative (dashed) vorticity are plotted along with 
the surface density (dotted).

For all runs, $a^2$ = 0.025, $\Gamma$ = 0.1, 
$\omega_0$ = 1 = 0.927 $\omega_{crit}$.  
\label{swirl}}
\end{figure}
\addtocounter{figure}{-1}
\begin{figure}[p]
\centerline{\resizebox{\colw}{\colw}{\includegraphics{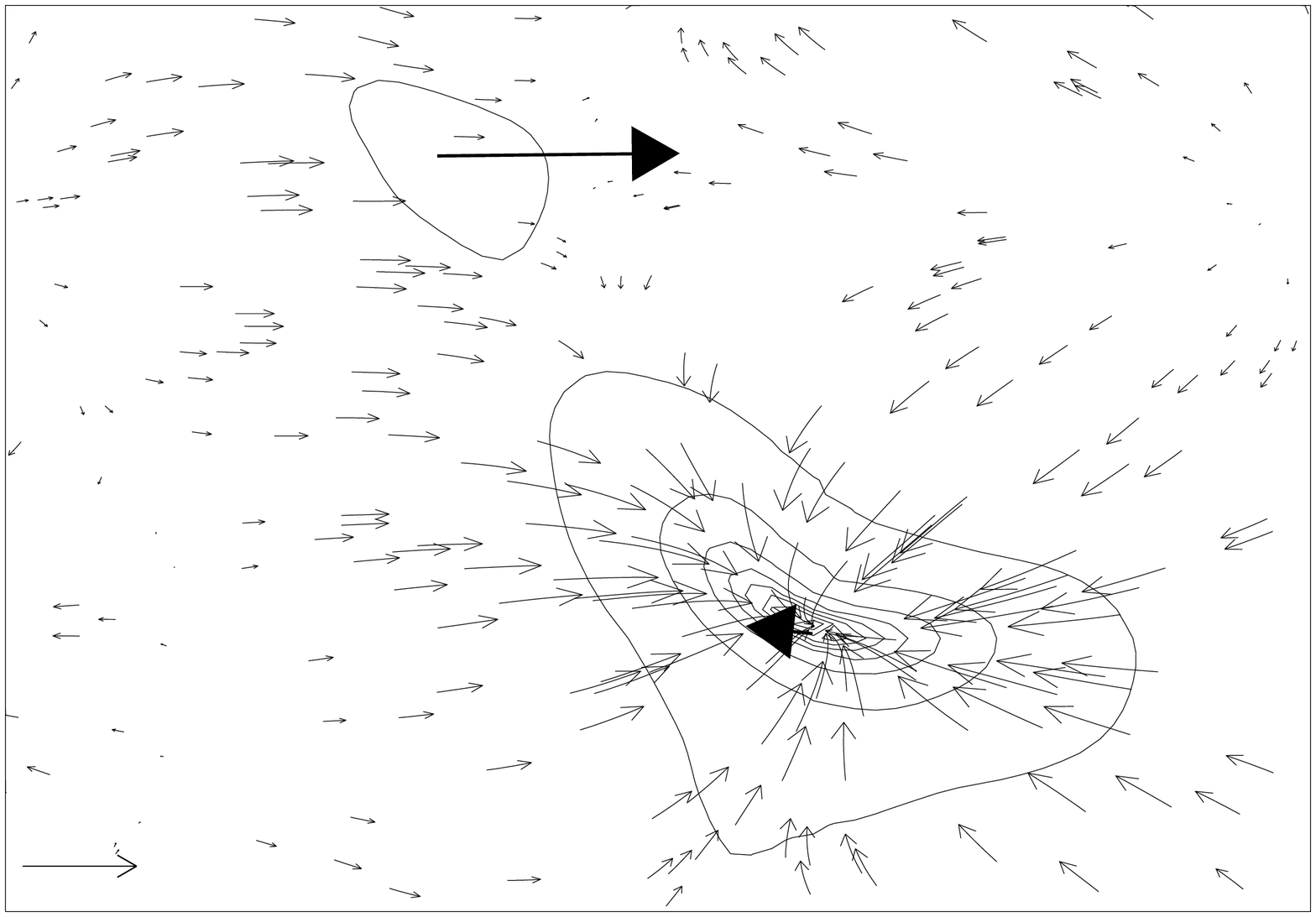}}
	\resizebox{\colw}{\colw}{\includegraphics{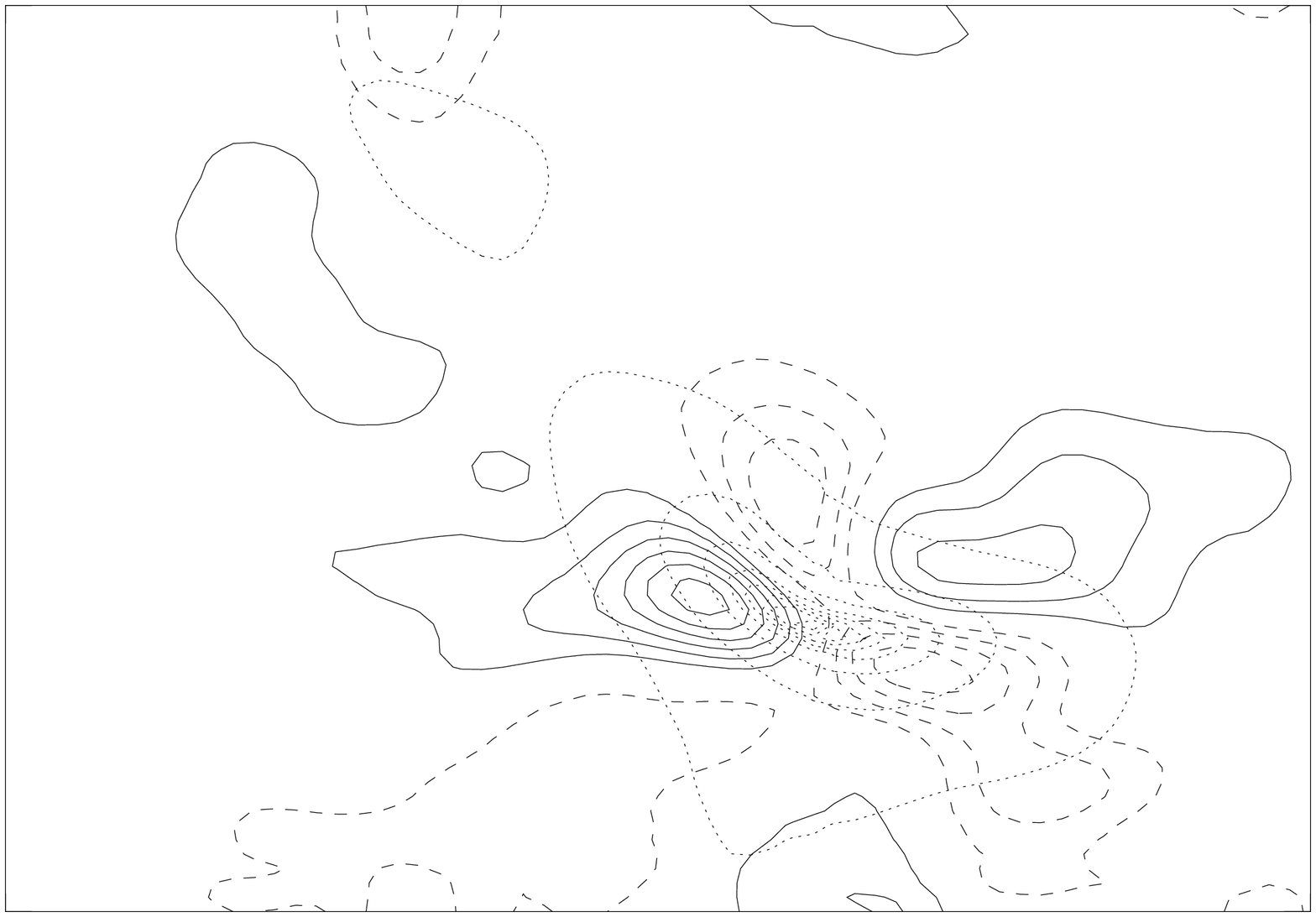}}}
\centerline{\resizebox{\colw}{\colw}{\includegraphics{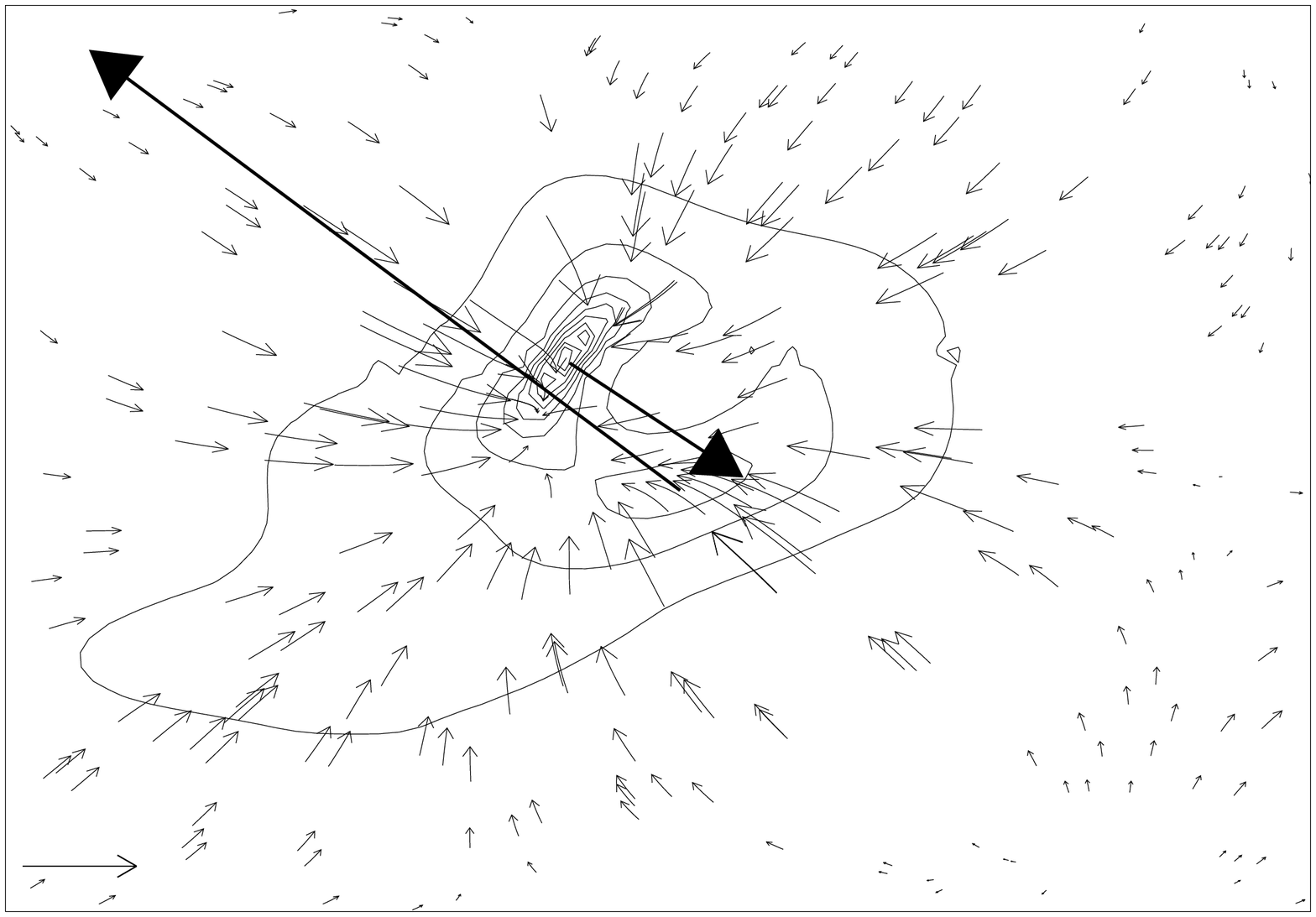}}
	\resizebox{\colw}{\colw}{\includegraphics{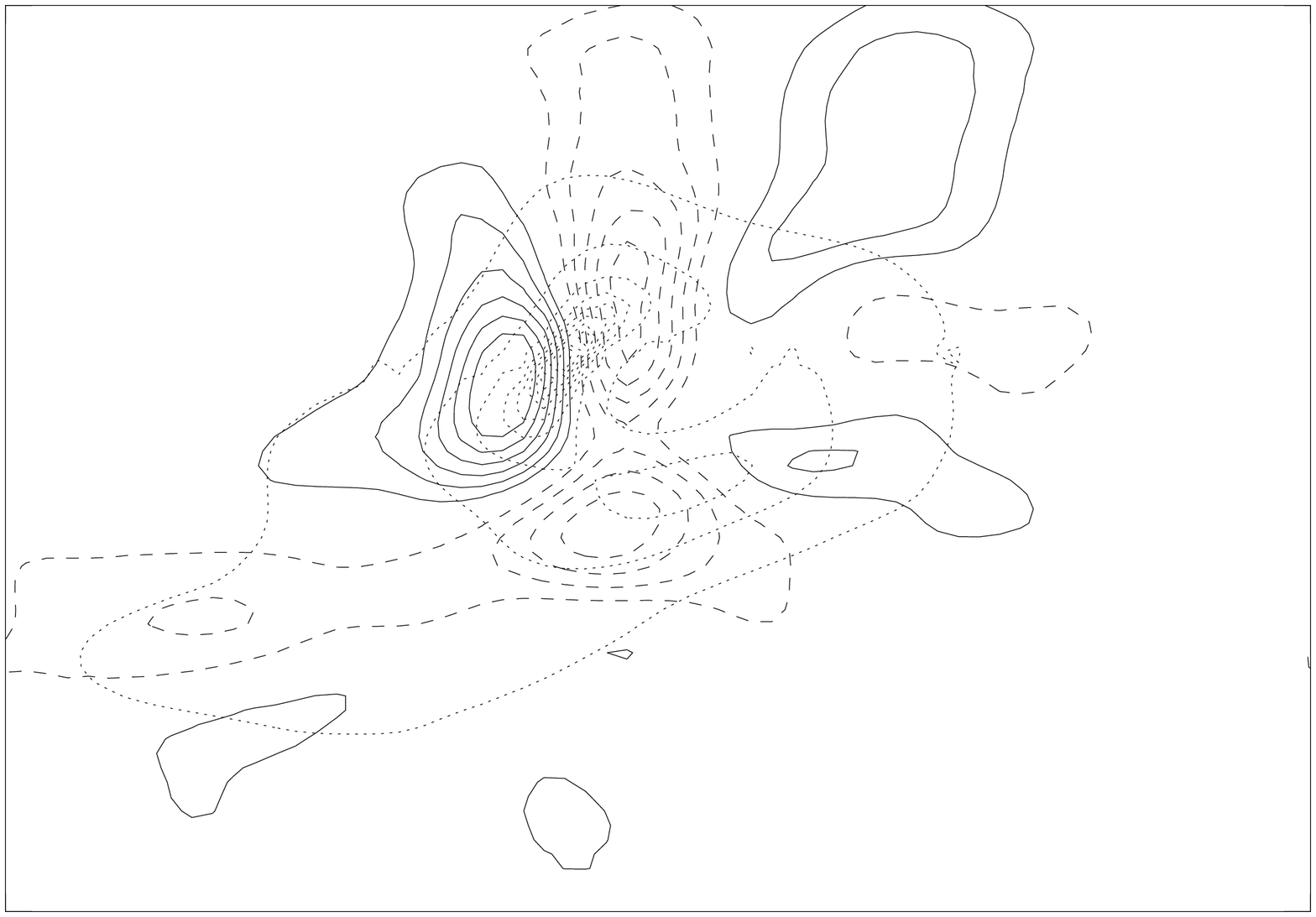}}}
\caption[Velocity and Vorticity]{continued.}
\end{figure}
\clearpage
\twocolumn

Visual inspection of Figure \ref{swirl} 
shows that the clumps are distinctly noncircular
and quite elongated in shape. Accounting for the third dimension, our clumps
should be considered triaxial or prolate. These shapes are consistent with
the measured and inferred shapes reported by \citet{m91}, \citet{r96}, and
\citet{wt99}.

\begin{figure}[hbt]
\centerline{\resizebox{\colw}{!}{\includegraphics{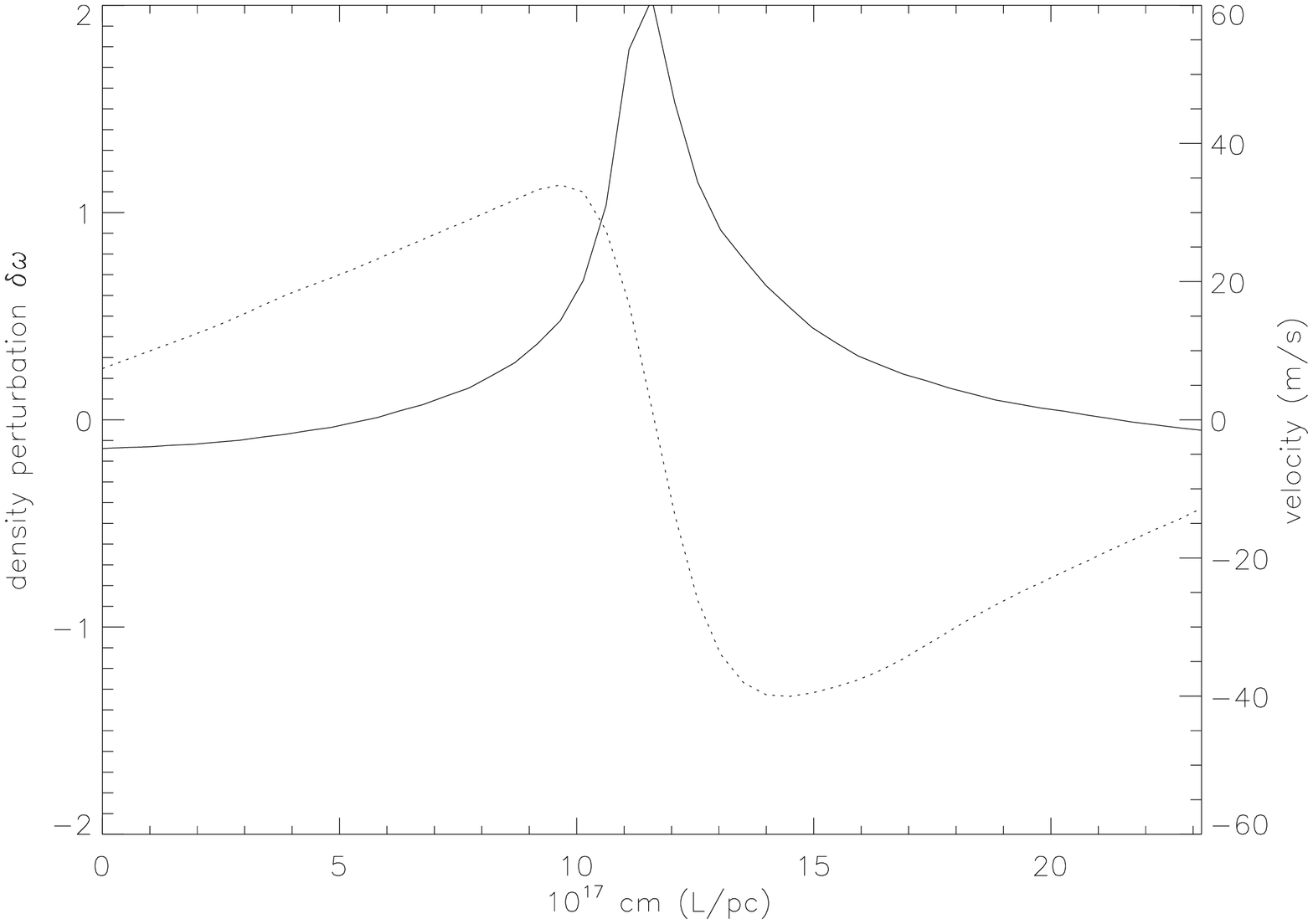}}}
\caption[]{
\renewcommand{\baselinestretch}{1}
\small\normalsize
Cross section of density perturbation $\delta\omega$ (solid)
and infall velocity (dotted) across a clump.  Infall velocity
increases towards the center of the clump (see text).  This run is
the third example in Figure \ref{swirl} - $\Gamma$ = 0.1, $a^2$ = 0.025,
$\omega_0$ = 1. = 0.927 $\omega_{crit}$. The collocation points are separated
by about 5 $\times$ 10$^{16}$ cm in this run, so the density and velocity
profiles are quite well resolved, except at the very center of the density
peak.
\label{cross}}
\end{figure}

Figure \ref{cross} shows a cross section of the surface density and velocity
profiles across the short axis of one particular collapsing clump. The density
is much more peaked than the velocity at this early stage of collapse; the
FWHM of the infall speed is several times larger than the FWHM of the density
peak. Both the velocity and density profiles are clearly, but not grossly,
asymmetric. 
Although it is premature to compare these density and velocity profiles
with observations of infall, it is encouraging that we see evidence for
extended inward motions as have been reported 
\citep{t98,greg,evans99,meo99}. A general feature of infall onto a line mass
such as a filament or strongly prolate object is that the velocity decays
more slowly with distance from the mass centroid than for infall onto a
spherically symmetric,
centrally concentrated object. This may be the main effect that produces the
extended infall.

The vorticity generated in swirling motions is $\nabla\times v\sim$
5/$t_{c0}\sim$ 8/Myr.  Collapse proceeds on a timescale of 
several Myr, so the swirling and 
rotation of clumps is not insignificant, although it is dominated by infall 
and we have not found evidence for clumps
torn apart by shear. Magnetic braking, which is excluded from our
calculation by the potential field approximation, would reduce rotation.
We estimate the magnetic braking rate in \S 4.1.  

Figure \ref{swirl} shows that the vorticity maxima are displaced from the
density maxima. In order to understand this, we derive an evolution
equation for the $\hat z$ component of vorticity, $\omega_z$, by taking the
curl of the equation of motion (\ref{gov2})
\begin{equation}
\label{vorticity}
{{\partial\omega_z}\over {\partial t}}+\boldsymbol\nabla_h\cdot(\omega_z
\mathbf{v}_h)=\mathbf{B}_h\times\boldsymbol\nabla_h {{B_z}\over {2\pi\sigma}}.
\end{equation}

According to equation (\ref{vorticity}), the generation of vorticity is second 
order in the amplitude of the fluctuation. There is generation of vorticity to 
first order only if there is a zero-order inclined field (Z98). We can
understand the spatial pattern of vorticity as follows. Despite
ambipolar drift, the contours of
constant $B_z$ track the contours of constant $\sigma$ quite well. Therefore,
the gradient of $B_z/\sigma$ is maximized toward the outer edge of a clump,
not at its center. Note that if the magnetic flux were perfectly frozen to the
matter, $B_z/\sigma$ would retain its initial constant value and there
would be no vorticity production at all. However, real clouds probably have
spatially varying $B_z/\sigma$, so in general, vorticity
production  does not require ambipolar drift. 
The maximum of $\mathbf{B}_h$, like the maximum gradient of $B_z/\sigma$,
 is displaced from the
clump center. Equation (\ref{vorticity}) shows that 
therefore vorticity is generated
off-center as well, and generally changes sign across the clump, so that clumps
are associated with vortex pairs. Moreover, equation (\ref{vorticity}) 
shows that
an axisymmetric clump does not generate vorticity. The dynamical pressure
of the vortices accentuates the non-axisymmetric nature of clump contraction,
and appears as streaming motions along the major axis of the clump. Although in
principle equation (\ref{vorticity}) 
suggests that the vortical velocity is scaled
by the Alfv\'{e}n speed, in
our numerical models the vortical
velocities are rather small, somewhat less
than the infall velocities. This is
large enough to noticeably elongate the clumps, but not enough to tear them 
apart by shear. 
%%%%%%%%%%%%%%%%%%%%%%%%%%%%%%%%%%%%%%%%%%%%%%%%%%%%%%%%%%%%%%%%%%%%%%%
\subsection{Energy Redistribution}\label{energy}

\begin{figure}[hbtp]
\centerline{\resizebox{\colw}{!}{\includegraphics{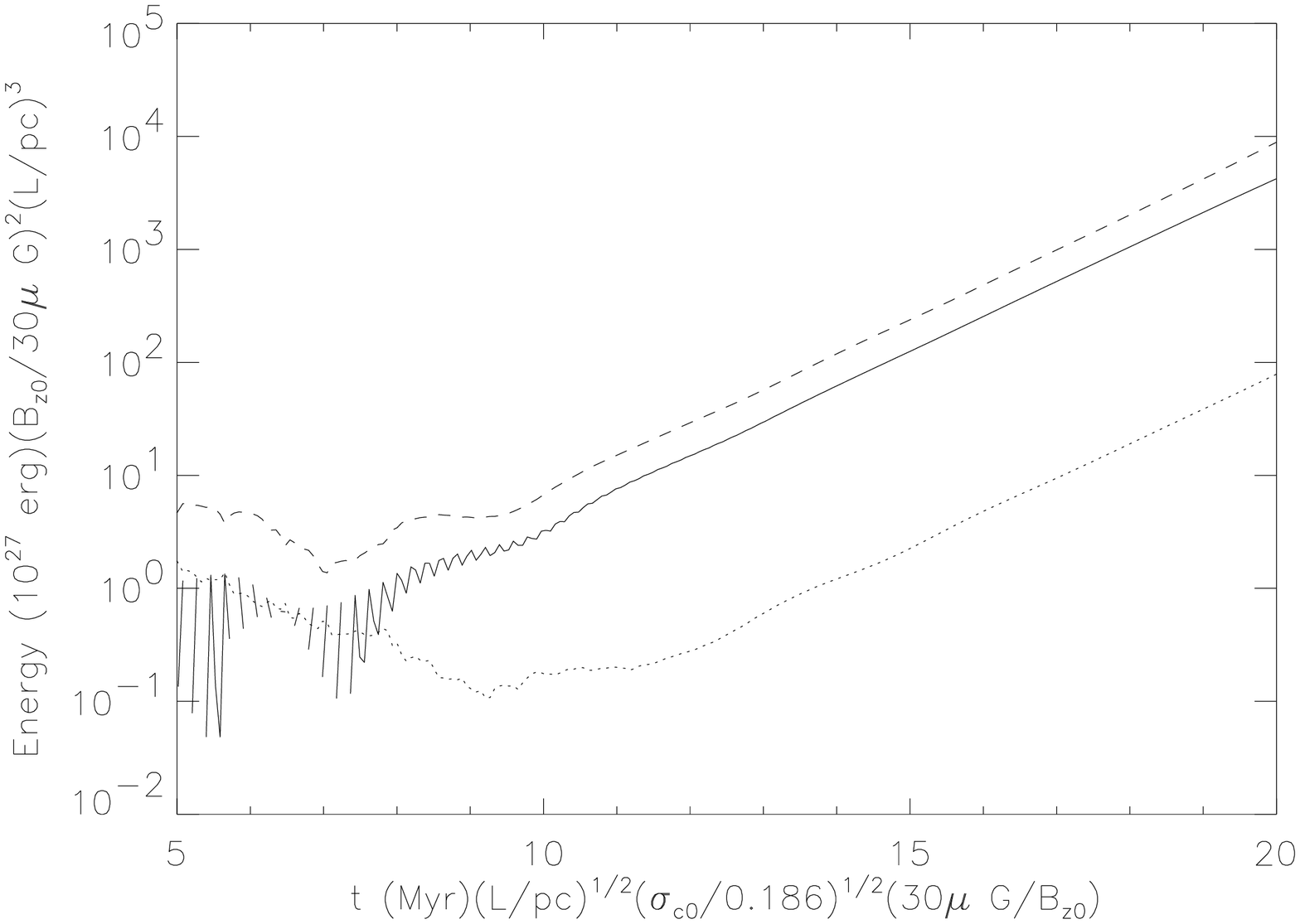}}}
\caption{\label{efig} Total energies in the simulation domain.  
Magnetic (dashed) and kinetic (dotted) energies are plotted 
with the absolute value of the gravitational (solid) energy.
The absolute values of all forms of energy increase exponentially
as collapse progresses, but the relative distribution does not 
change significantly (see text.) For this particular run, 
$\Gamma$ = 0.1, $\omega_0$ = 0.973 = 0.864 $\omega_{crit}$, 
and $a^2$ = 0.04.
}\end{figure}

Z98 suggested that this magneto-gravitational instability 
might generate significant turbulent kinetic energy
by releasing energy contained in the background magnetic field. 
Analysis of the total gravitational, magnetic, and kinetic 
energy (Fig. \ref{efig})
in these simulations shows that the absolute values of 
all three forms of 
energy grow exponentially during collapse.  The magnetic 
energy, which is the fluctuation energy integrated over the space outside the
disc, dominates at all stages of collapse in these clouds, 
which would be stabilized by the magnetic field in the absence 
of ambipolar drift.  
%Kinetic energy is generated along with 
%the significant vorticity discussed above, but
%there does not appear to be a preferential generation of kinetic 
%energy or turbulent motions. 
The initial magnetic field is uniform in these
simulations, and so there is no stored magnetic energy available for
conversion to turbulent motions. The relatively low kinetic energy in the
models is a signature of the importance of diffusive, as opposed to dynamical,
effects.
In ideal MHD turbulence with self gravity one would expect 
equipartition between the kinetic and potential energies \citep{zm95}. In this
case the potential energy is the sum of the gravitational and magnetic 
energies, but Figure 9 shows that the kinetic energy is about an order of 
magnitude less than the equipartition value.

%%%%%%%%%%%%%%%%%%%%%%%%%%%%%%%%%%%%%%%%%%%%%%%%%%%%%%%%%%%%%%%%%%%%%%
\section{Discussion of Approximations}\label{discuss}

It is difficult and computationally
expensive to simulate the evolution of magnetized,
self gravitating molecular clouds in three dimensions with sufficient 
resolution to capture all the relevant physical processes. 
In this study we focussed on the two-dimensional
instabilities of a sheetlike cloud surrounded by a conducting medium without
inertia. This allowed us to approximate the external magnetic field as current
free, and to work in only two spatial dimensions. In the next two 
subsections we discuss the accuracy of these approximations.
 
\subsection{Potential Field Approximation}\label{potfield}

In the Appendix, we show how to extend each Fourier component of the
magnetic and velocity 
perturbations above and below the sheet. Although the perturbations become 
highly nonlinear within the cloud, nonlinear effects outside the cloud 
are weak as long as the velocities inside the cloud are sub-Alfv\'{e}nic 
with respect to the intercloud medium, which is expected for low ambient
density.
The compressive part of the cloud velocity field
generates evanescent, fast magnetosonic waves which decay exponentially
away from the disc, while the vortical part generates Alfv\'{e}n waves which
propagate away from the disc. 

Both the magnetosonic and Alfv\'{e}n waves slightly change the horizontal 
magnetic field perturbation in the disc, thereby changing the magnetic 
force from its value in the potential approximation. If we define
an external Alfv\'{e}n timescale $\tau_{Ae}\equiv (kv_{Ae})^{-1}$ 
for wavenumber 
$k$ in the disc, and let the timescale for the perturbation in the 
cloud be $\tau_c$, then according to equation (\ref{A9}) the correction to 
the force due to compressive motion is of order $(\tau_{Ae}/\tau_c)^2$, 
while the correction due to vortical motion is of order 
$(\tau_{Ae}/\tau_c){\cal C}_0/{\cal D}_0$, where ${\cal C}_0/{\cal D}_0$ 
is the ratio of the amplitude of compressive to noncompressive motion. 
If the motions in the disc were Alfv\'{e}nic, $\tau_{Ae}/\tau_c$ 
would be of order the ratio of the cloud density to external density 
$(\rho_c/\rho_e)^{1/2}$, but the perturbation
frequency is sub-Alfv\'{e}nic, so the ratio of timescales is even larger. 
Moreover, the perturbations are primarily compressive rather than vortical. 
Thus, the error in the force incurred by the potential approximation is 
likely to be small. Even if the density contrast were only 10$^2$, and the
motions in the disk were Alfv\'{e}nic and purely vortical, the 
potential field approximation would still be accurate to 10\%.

The Alfv\'{e}n wave flux tends to suppress the instability, and removes 
vorticity from the cloud, at a rate that we can quantify. We define 
an energy damping time $\gamma_d$ as the ratio of outward propagating 
energy flux to the vertically integrated wave energy in the disc. 
From equation (\ref{A10}),
\begin{equation}
\label{eqn_damp}
\gamma_d = {{2\rho_ev_{Ae}}\over{\sigma}}
	{{\mid{\cal C}_0^2\mid}\over{(\mid{\cal C}_0^2\mid + 
	\mid{\cal D}_0^2\mid)}}.
\end{equation}

If we replace $\sigma$ by the critical surface density $B_{0z}/2\pi G^{1/2}$ 
and assume that the vertical fields inside and outside the cloud are the 
same then $\gamma_d$ is just the gravitational frequency for the intercloud 
medium, reduced by the ratio of vortical to total kinetic energy
\begin{equation}
\gamma_d = (4\pi G\rho_e)^{1/2}{{\mid{\cal C}_0^2\mid}\over {(\mid{\cal
C}_0^2\mid + \mid{\cal D}_0^2\mid)}}.
\end{equation}

Since self gravity is presumably negligible in the low density 
intercloud medium, the energy loss rate is negligible as well. 
Loss of vorticity is measured by the magnetic braking rate 
$\gamma_{mb}$, which can be shown by a similar argument to be
$\gamma_{mb} = (4\pi G\rho_e)^{1/2}$.

We thus see how the potential field limit is approached as the density contrast
between the cloud and intercloud medium increases. At the late stages of clump
formation the potential field becomes highly distorted and develops partially
closed topology, but we can ignore this for the relatively mild density
contrasts studied in this paper.

\subsection{Approximations to the Gas Physics}

The two dimensional approximation has a long and venerable history in galactic
dynamics and accretion disc theory, as well as in studies of molecular 
clouds, and its errors for self gravitating systems are reasonably well 
understood. We expect the approximation to be reasonably good
as long as the clump diameters exceed the disc thickness. The instability 
discussed in this paper 
has a 3D analog (Z98), but it must be treated by other means.

We assumed that the gas has an isothermal equation of state. This is a 
reasonable description of the kinetic pressure - density relationship, 
because of the high radiative efficiency of molecular gas. However, if 
the pressure were due to unresolved turbulence the medium would 
generally be less compressible; for example, Alfv\'{e}n wave pressure follows 
density according to a 3/2 law \citep{mz95}. This would make the medium 
more stable by increasing the value of $\gamma_T$ (see eq. [\ref{disp_rel}]),
as would retention of magnetic pressure.

We took a uniform sheet at rest as an initial condition. This has the 
advantage of simplicity, but it means that there is no free energy stored 
in the background magnetic field. Thus, we have not tested the conjecture 
that the instability can convert magnetic free energy to turbulent energy, 
which was proposed in Z98.

We have treated ambipolar drift in the strong coupling approximation, and have
implicitly assumed that $v_D <$ 20 km s$^{-1}$ (otherwise the rate coefficient
would change). This is reasonable as long as the ion-neutral collision time,
which is of order $5\times 10^9 n_n^{-1}$ s, is
less than other timescales in the problem.

We have parameterized the relationship between the collision rate and the
surface density by an exponent $\alpha$. In order to do better we would need a
three dimensional model of the sheet and might need to follow the ionization
as well. The results of linear theory are rather insensitive to the value of
$\alpha$, which suggests that it need not be calculated very accurately
in the present models.
%%%%%%%%%%%%%%%%%%%%%%%%%%%%%%%%%%%%%%%%%%%%%%%%%%%%%%%%%%%%%%%%%%%%%%%
\section{Summary}\label{summary}

The study of axisymmetric 
contraction of weakly ionized,
self gravitating, magnetized clouds has proceeded quite far
\citep{bm94,cm93,cm94,sms97,ck98}. In this paper,
our emphasis has been on the initial breakup of a cold magnetized cloud 
gas into fragments and the early stages  of their magnetic flux loss and
contraction.
We include self gravity,
magnetic tension, and ambipolar drift, but we do not include detailed
chemistry of grain physics, choosing instead a simple parameterization of the
ionization. We follow the
evolution for a shorter time than the isolated cloud collapse studies, but
we impose no symmetry constraints or initial density or velocity structure.

We study highly flattened clouds, with 
an initially perpendicular magnetic field, which are slightly subcritical.  
The linear theory of 
collapse in such geometry (Z98, T=0; this work, T$>$0) predicts
collapse on an intermediate timescale, faster than the 
diffusive timescale set by ambipolar drift, but slower than 
the dynamical timescale of free-falling inside-out collapse. 
The linear theory also predicts the existence of a single 
spatial wavenumber with maximal growth rate, with sufficiently short 
wavelengths stabilized by thermal pressure. This naturally suppresses power
at short wavelengths, which is important for the success of the spectral
method we employ in the simulations.

We simulate collapse in clouds with random initial density perturbations
which grow from $<$0.1\% of the mean density to 5-10 times the mean 
density.  We confirm the intermediate collapse rate predicted by 
linear theory (\S\ref{gamgam}), although the
nonlinear collapse rate is faster than the linear rate. These intermediate
rates are consistent with some recent observations of infall in molecular
clouds \citep{evans99,meo99}.
 
We show that clouds fragment into 
clumps with size corresponding to the wavelength of the spatial 
mode of maximal linear growth rate (\S\ref{clump}), generally 1-10 
$M_{\odot}$.  Collapse 
is asymmetric and complex (\S\ref{vel}), and
generally forms prolate clumps, for which there is observational evidence
\citep{m91,r96,wt99}. Sometimes the clumps are in
mutual orbit, although the typical clump separation, a few tenths of a parsec,
is too large to be relevant to the formation of binary stars. The magnetic
field drives the growth of local vorticity, typically in the form of vortex
pairs which straddle the clumps and are associated with streaming motions along
them.  

Considerable magnetic
flux is lost from the collapsing clumps, consistent with the 
observationally determined $|\mathbf{B}|\propto\rho^{0.5}$ 
\citep{troland86,c99} (see also \S\ref{brho}). 
This flux loss is consistent with other calculations of cloud 
evolution \citep{fm93,cm94}
(although ambipolar drift is not necessary to bring 
about this relationship, either for isolated \citep{mous76} or 
turbulent, highly structured \citep{pn99} clouds). As magnetic flux is lost
and the surface density increases in the central regions of a contracting
core, further fragmentation might ensue. 

One prediction of the linear theory, namely that the instability could 
convert magnetic free energy to turbulence, has not been borne out by 
the simulations. This may be due to the fact that the initial magnetic 
field is completely uniform and therefore carries no free energy. Although 
significant magnetic curvature develops late in the runs, the cloud has 
already become quite dynamical.  This prediction awaits future tests with 
a more stressed initial state. The relative motions of the clumps shown
in Figure 7 are about an order of magnitude less than the relative motions
of clumps separated by a few tenths of a parsec in real clouds \citep{g98}, 
although the infall velocities in the simulation are comparable to 
measured velocities \citep{meo99}.

An interesting area of future work would be to extend this 
study to true three-dimensional clouds. 
The linear theory (Z98) indicates that this
instability exists in three as well as two dimensions,
and that the growth rate is still intermediate to slow 
diffusive contraction and fast dynamical collapse.  
Construction of a nonlinear model in three dimensions would be more difficult
than the two dimensional models developed here,
but could prove interesting.  In this vein, we find the recent successful
fit of observations of L1544 with a nearly critical model
\citep{cb2000} encouraging.

\acknowledgements
We are happy to acknowledge support by NSF grant AST 9800616, 
a 3-year NSF Graduate Research Fellowship to R.I., and NASA 
grant NAG 5-4063 to the University of Colorado, as well as discussions with
Neal Evans and comments by an anonymous referee.

%%%%%%%%%%%%%%%%%%%%%%%%%%%%%%%%%%%%%%%%%%%%%%%%%%%%%%%%%%%%%%%%%%%%%%
%\renewcommand{\baselinestretch}{1.}

%%%%%%%%%%%%%%%%%%%%%%%%%%%%%%%%%%%%%%%%%%%%%%%%%%%%%%%%%%%%%%%%%%%%%%
\appendix
\section{Appendix}

In this Appendix we drop the potential field approximation and calculate the
response of the intercloud medium to motions within the cloud. This allows us
to estimate the errors incurred by assuming a potential field.

We carry out the estimate using linearized, ideal MHD theory. This
is more accurate in the intercloud medium than it would be
in the disc, because the intercloud or external Alfv\'{e}n 
speed $v_{Ae}$ is relatively high while
ion-neutral friction is weak. We assume the equilibrium intercloud
field is uniform and vertical (${\bf B_0}=\hat z B_0$). We
choose to work in the half space $z>0$ (the results are similar in
the other half space). In linear
theory, the motions are purely horizontal, and we can derive the
following pair of decoupled equations for the divergence ${\cal D}$
and vertical component ${\cal C}$ of the curl of the velocity
\begin{mathletters}
\begin{eqnarray}
\Bigg({{\partial^2}\over {\partial t^2}} - v_{Ae}^2\nabla^2\Bigg)
{\cal D} &=& 0,\label{A1} \\
\Bigg({{\partial^2}\over {\partial t^2}} - v_{Ae}^2
{{\partial^2}\over {\partial z^2}}\Bigg){\cal C}&=&0\label{A2}.
\end{eqnarray}
\end{mathletters}
Equations (\ref{A1}) and (\ref{A2}) represent fast magnetosonic waves and
Alfv\'{e}n waves, respectively. In general, both types of waves are
generated by the motions in the cloud.

In order to make progress, we assume plane wave horizontal behavior
and exponential behavior in time, so that all perturbations 
depend on ($x$, $y$, $t$) as
$\exp(\gamma t + ik_x x + ik_y y)$, where $\gamma$ may be
complex: $\gamma = i\omega + \nu$ with both $\omega$, $\nu > 0$. Then
\begin{equation}
{\cal C}=i(k_xv_y-k_yv_x);\quad {\cal D}=i(k_xv_x+k_yv_y)
\end{equation}
can be calculated at $z=0$ in terms of the motions on the disc.
The vertical extensions of these quantities can be found from 
equations (\ref{A1}) and (\ref{A2}), choosing outward
going or exponentially decaying wave solutions. For the Alfv\'{e}nic part,
\begin{equation}
{\cal C}={\cal C}_0e^{-ik_Az};\quad k_A\equiv{{\gamma}\over {v_{Ae}
}},\label{A3}
\end{equation}
where ${\cal C}_0$ is the value of ${\cal C}$ at $z=0$.
For the magnetosonic part,
\begin{equation}
{\cal D}={\cal D}_0e^{-k_Mz};\quad k_M\equiv k_{\perp}\Bigg(1+
{{\gamma^2}\over {k_{\perp}^2v_{Ae}^2}}\Bigg)^{1/2} \label{A4},
\end{equation}
where $k_{\perp}^2\equiv k_x^2 + k_y^2$.
In equation (\ref{A3}) we have written the vertical dependence as a propagating
wave, and in equation (\ref{A4}) as an evanescent wave. Although both $k_A$
and $k_M$ are complex, because $\gamma$
is complex, our notation reflects the salient aspects
of their behavior. The magnetosonic wave is almost purely evanescent
because the wave frequency is much less than the disc Alfv\'{e}n 
frequency, which in turn is much less than the intercloud Alfv\'{e}n
frequency. The Alfv\'{e}n wave has a substantial propagating component
and decays in space as long as the disturbance is growing in time,
which is purely a result of causality.

We now calculate the perturbed magnetic field components
${\bf\delta B}$ at the
disc. According to the linearized induction equation,
\begin{mathletters}
\begin{eqnarray}
{{\partial{\bf\delta B}_{\perp}}\over {\partial t}} &=& B_0
{{\partial{\bf v}_{\perp}}\over {\partial z}},\label{A5}\\
{{\partial\delta B_z}\over {\partial t}} &=& -B_0{\cal D}.\label{A6}
\end{eqnarray}
\end{mathletters}
Inverting the definitions of ${\cal C}$ and ${\cal D}$ for the
velocity components gives
\begin{equation}
v_x={{i}\over {k_{\perp}^2}}(k_y{\cal C}-k_x{\cal D});\quad v_y =
-{{i}\over {k_{\perp}^2}}(k_x{\cal C}+k_y{\cal D}).\label{A7}
\end{equation}
Using equations (\ref{A3}), (\ref{A4}), and (\ref{A7}) 
in equations (\ref{A5}) and (\ref{A6}) 
gives the field components at $z=0$
\begin{mathletters}
\begin{eqnarray}
\delta B_x &=&{{B_0}\over {k_{\perp}^2\gamma}}(k_Ak_y{\cal C}_0 +
ik_Mk_x{\cal D}_{0}),\label{A8a} \\
\delta B_y &=&{{B_0}\over {k_{\perp}^2\gamma}}(-k_Ak_x{\cal C}_0 +
ik_Mk_y{\cal D}_{0}),\label{A8b}\\
\delta B_z&=&-{{B_0}\over {\gamma}}{\cal D}_0.\label{A8c}
\end{eqnarray}
\end{mathletters}

We can use equations (\ref{A8a}-\ref{A8c}) 
to compare the MHD solution with the potential
field limit. The ratios of the perturbed horizontal to vertical
field components at $z=0$ can be written as
\begin{equation}
{{\delta{\bf B}_{\perp}}\over {\delta B_z}}=-i\hat k_{\perp}\Bigg(1
+{{\gamma^2}\over {k_{\perp}^2 v_{Ae}^2}}\Bigg)^{1/2} + (\hat z\times
\hat k_{\perp}){{\gamma}\over {k_{\perp}v_{Ae}}}{{{\cal C}_0}\over {{\cal
D}_0}}.\label{A9}
\end{equation}
Equation (\ref{A9}), together with equation (\ref{A3}), 
shows that in the limit
$v_{Ae}\rightarrow\infty$ the potential field solution is exact.

The Alfv\'{e}nic part of the disturbance, as a propagating wave, removes
both energy and angular momentum from the cloud. The energy flux
${\cal F}_W$
(accounting for both kinetic and electromagnetic energy, and for
waves propagating in both directions away from the disc) is
\begin{equation}
{\cal F}_W=2\rho_e{{\mid{\cal C}_0^2\mid}\over {k_{\perp}^2}}v_{Ae}.
\label{A10}
\end{equation}
In \S\ref{potfield} we use equation (\ref{eqn_damp}) 
to derive the rate at which the perturbation in the
disc is damped by outgoing waves.

\end{document}